\documentclass{PoS}

\usepackage{overpic}
\usepackage{axodraw}
\usepackage{axodraw}
\newcommand{\be}{\begin{equation}}
\newcommand{\ee}{\end{equation}}
\newcommand{\ba}{\begin{eqnarray}}
\newcommand{\ea}{\end{eqnarray}}

\usepackage{marvosym}
\title{Quark Mass dependence at Two Loops for Meson Properties}

\ShortTitle{Quark Mass dependence at Two Loops for Meson Properties}

\author{\speaker{Johan Bijnens}%
        \\
        Lund University\\
        E-mail: \email{bijnens@thep.lu.se}}

\abstract{This talks contains a short introduction to
Chiral Perturbation Theory and the existing calculations to two-loop order
in the mesonic sector. I include a discussion on which quantities
the expansion can be organized in.
The present best values of the
Low-Energy-Constants
as determined from continuum physics are given as well as the assumptions
underlying the fits to experimental data.
I present plots of masses, decay constants and $f_+(0)$ in $K_{\ell3}$
as a function
of quark or meson masses.
The talk ends with a list of things for which it would be extremely useful to
have good results from lattice QCD calculations.}

\FullConference{The XXV International Symposium on Lattice Field Theory\\
		 July 30 - August 4 2007\\
		 Regensburg, Germany}

\renewcommand{\logo}
{\raisebox{-5mm}
{\parbox{\textwidth}{\begin{flushright}LU TP 07-22\\
arXiv:0708.1377v2 [hep-lat]\\
revised August 2007
\end{flushright}
\includegraphics{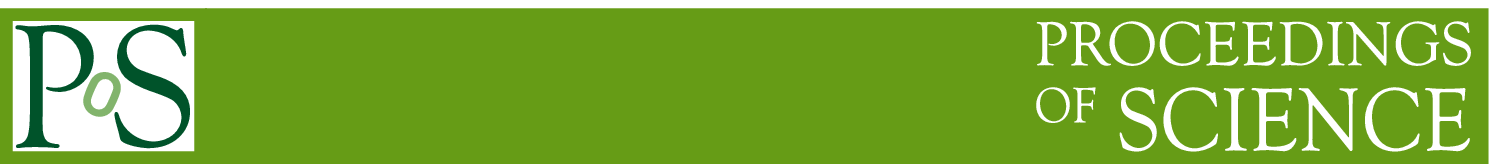}}
}}
\FullConference{{\rm To be published in the proceedings of:\\}
The XXV International Symposium on Lattice Field Theory\\
		 July 30 - August 4 2007\\
		 Regensburg, Germany}

\begin{document}

\section{Introduction}

An alternative title for this talk is\\[2mm]
\centerline{
\framebox{\parbox{0.95\textwidth}{
What is Known About Low Energy Constants and Quark Mass
Dependence in Chiral Perturbation Theory \emph{from the Continuum}}}}\\[2mm]
In order to discuss this a short introduction to Chiral perturbation Theory
(ChPT) is
given with an emphasis on some of the aspects that have been known to create
confusion, the choice of quantities to express the expansion in.
The remainder can be split in an overview of the existing calculations
in two and three-flavour ChPT and quark mass dependences of several of
the quantities of interest in both cases. I restrict myself here to the
cases where a full two-loop calculation is available and to quantities of
interest to the lattice community.
I also remind you of the
existing partially quenched calculations at two-loop order.
In the remainder I will use order $p^6$ in the chiral
counting, next-to-next-to-leading order (NNLO) and two-loop order
as synonyms even though strictly speaking the two-loop diagrams are only
part of the NNLO result.

A review of order $p^6$ ChPT is \cite{reviewp6} and several more references
to lectures as well as files containing the long two-loop expressions
can be found in the website \cite{website}. I want to point out those by Sharpe
aimed at lattice QCD practitioners \cite{Sharpelectures}.

\section{Chiral Perturbation Theory}

\emph{Chiral perturbation Theory} in its modern form was introduced by
Gasser, Leutwyler and Weinberg \cite{Weinberg0,GL0,GL1} and it can defined
as
\\[2mm]\centerline{
\framebox{\parbox{0.95\textwidth}{Exploring the consequences of the
chiral symmetry of QCD
and its spontaneous breaking using
effective field theory techniques}}}\\[2mm]
The assumptions that are needed to allow a derivation from QCD and a full
derivation can be found in the paper by Leutwyler \cite{Leutwyler1}.
I do not intend to give a full derivation here but only restrict myself to
a few comments.

\emph{Powercounting:} In effective field theory, one assumes that there is a gap
in the spectrum which allows to include only the degrees of freedom that
are relevant below the gap and treat the effects of the degrees of freedom
at higher scales perturbatively. Thus a clear separation of scales is the
first requirement. One then constructs the most general local Lagrangian with
the lower degrees of freedom in agreement with the symmetries and their
realization. Unfortunately, this leads to an infinite
number of parameters and hence there is no predictivity left. If there
exists a way to organize the series in terms of order of importance,
then we can work order by order and have predictivity at any fixed order
in the importance. This is typically achieved by introducing
\emph{powercounting}.

\emph{Unitarity:} In Ref.~\cite{Leutwyler1} one uses strongly the fact that
the only singularities at low energies
in Green functions come from poles and cuts of the light degrees of freedom
and that all the remaining vertices can be expanded. This is where unitarity
plays a role in the derivation of ChPT from QCD.

\emph{ChPT:} ChPT is the effective field theory build with the Goldstone
Bosons resulting from the spontaneous breakdown of chiral symmetry in QCD
as primary degrees of freedom. The powercounting used\footnote{There
are different countings possible, the one here is the standard one.
References can be found in \cite{reviewp6} and the talk
by S. Descotes-Genon\cite{Descotes-Genon}.}
is basically dimensional counting in momenta and (meson) masses.
The expected breakdown scale is the scale of resonances, so for energies
around $m_\rho$ standard mesonic ChPT definitely does not work.
The breakdown scale depnds a bit on the channel.

\emph{Chiral Symmetry:} In QCD with three light quarks
of equal mass, they are fully interchangeable and we have
a $SU(3)_V$ symmetry.
But looking at the QCD Lagrangian
\be
{\cal L}_{QCD} =  \sum_{q=u,d,s}
\left[i \bar q_L D\hskip-1.3ex/\, q_L +i \bar q_R D\hskip-1.3ex/\, q_R
- m_q\left(\bar q_R q_L + \bar q_L q_R \right)
\right]
\ee
we see that for  $m_q = 0$ we have a separate interchange for
the left and right-handed quarks: $G_\chi= SU(3)_L \times SU(3)_R$.

\begin{figure}[ht]
\vskip-2cm
\hskip1cm
\begin{minipage}{0.3\textwidth}
\unitlength=0.5pt
\begin{picture}(100,100)
\SetScale{0.5}
\SetWidth{1.5}
\Line(0,100)(100,0)
\Line(0,0)(100,100)
\Vertex(50,50){5}
\end{picture}
\hfill\raisebox{25pt}{$p^2$}\\[0.25cm]
\unitlength=0.5pt
\begin{picture}(100,30)
\SetScale{0.5}
\SetWidth{1.5}
\Line(0,15)(100,15)
\end{picture}
\hfill\raisebox{5pt}{$1/p^2$}\\[0.25cm]
$\int d^4p$\hfill$p^4$
\end{minipage}
\hfill
\raisebox{0.75cm}{
\begin{minipage}{0.45\textwidth}
\begin{picture}(100,100)
\SetScale{0.5}
\SetWidth{1.5}
\Line(0,100)(20,50)
\Line(0,0)(20,50)
\Vertex(20,50){5}
\CArc(50,50)(30,0,180)
\CArc(50,50)(30,180,360)
\Vertex(80,50){5}
\Line(80,50)(100,100)
\Line(80,50)(100,0)
\end{picture}
\hskip-1cm\raisebox{25pt}
{$(p^2)^2\,(1/p^2)^2\,p^4 = p^4$}\\[0.25cm]
\unitlength=0.5pt
\begin{picture}(100,100)
\SetScale{0.5}
\SetWidth{1.5}
\Line(0,0)(50,40)
\Line(0,50)(50,40)
\CArc(50,70)(30,0,180)
\CArc(50,70)(30,180,360)
\Vertex(50,40){5}
\Line(50,40)(100,50)
\Line(50,40)(100,0)
\end{picture}
~~\raisebox{25pt}
{$(p^2)\,(1/p^2)\,p^4 = p^4$}
\end{minipage}
}
\caption{\label{figpower} An illustration of the power-counting in ChPT.
On the left we see the lowest order vertex with two powers of momenta
or masses,
the meson propagator with two inverse powers and the loop integration leading to four
powers. On the right hand-side we see two one-loop contributions and how the
counting on the left leads to the same power $p^4$ for both diagrams.
This counting was generalized to all orders in \cite{Weinberg0}.}
\end{figure}
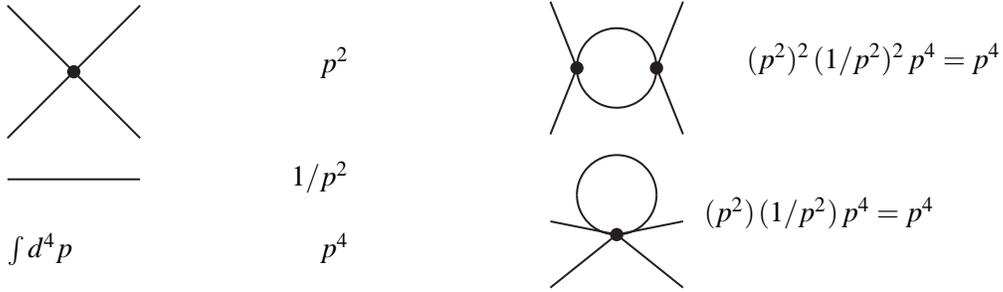

\emph{Chiral Symmetry Breaking:} The fact that the vacuum expectation value
$
\langle \bar q q\rangle = \langle \bar q_L q_R+\bar q_R q_L\rangle
\ne 0
$ leads to the spontaneous breaking of
$ SU(3)_L \times SU(3)_R$ to
$SU(3)_V$.
The eight broken generators lead to eight massless degrees of freedom
\emph{and} makes their interactions vanish at zero momentum.
The latter fact allows to produce a consistent powercounting in ChPT.
This is illustrated in Fig.~\ref{figpower}.

\emph{Chiral Perturbation Theor{\bf ies\,}:} ChPT is a very large subject,
more than 3500 papers cite at least one of the three basic papers. The
name ChPT also is given to a variety of different theories exploiting
the chiral symmetry of QCD. These include application to baryons, 
mesons and baryons containing heavy quarks, vector mesons and other resonances
and of course the light pseudoscalar mesons. Within the latter we can
distinguish between theories with two, three or more flavours, also in the
partially quenched varieties, as well as including electromagnetism
and the weak interactions nonleptonically and the possibility of treating
the kaon as a heavy particle. This talk restricts itself to the standard
two, three or more flavour sector with strong interactions and couplings to
external currents. This is the part that has been most fully pushed to
order $p^6$.

\emph{Lowest order Lagrangian:} The Goldstone Bosons live on the manifold $G/H$
with $G=SU(n_F)_L\times SU(n_F)_R$ and $H=SU(n_F)_V$ for $n_F$ flavours of
quarks. $G/H$ has again the
structure of $SU(n_F)$ and can be parametrized by a matrix
\be
 U(\phi) = \exp(i \sqrt{2} \Phi/F_0),
\quad\mbox{with}\quad
\Phi (x) = \,
{ \left( \begin{array}{ccc}
\frac{ \pi^0}{ \sqrt 2} \, + \, \frac{ \eta_8}{ \sqrt 6}
 & \pi^+ & K^+ \\
\pi^- & - \frac{\pi^0}{\sqrt 2} \, + \, \frac{ \eta_8}
{\sqrt 6}    & K^0 \\
K^- & \bar K^0 & - \frac{ 2 \, \eta_8}{\sqrt 6}
\end{array}  \right)} .
\ee
Here I have indicated the traceless matrix $\Phi$ in terms of the
known pseudoscalars for the case of $n_F=3$.
The lowest order or order $p^2$ Lagrangian contains two terms
\be
{\cal L}_2 = \left({F_0^2}/{4}\right)
\left\{\langle D_\mu U^\dagger D^\mu U \rangle 
+\langle \chi^\dagger U+\chi U^\dagger \rangle \right\}\, ,
\ee
with a covariant derivative
$
D_\mu U = \partial_\mu U -i r_\mu U + i U l_\mu \,,
$
which includes the
left and right external currents: $
r(l)_\mu = v_\mu +(-) a_\mu$.
The scalar and pseudoscalar external densities are included via
$\chi = 2 B_0 (s+ip)$ with the quark masses included in the
 scalar density: $s= {\cal M} + \cdots$. The notation
$\langle A \rangle$ is the trace over flavours $Tr_F\left(A\right)$.

\emph{NLO Lagrangian:} The Lagrangian at order $p^4$ was classified in \cite{GL0,GL1} and is for $n_F=3$
\ba
{\cal L}_4 &=&
L_1 \langle D_\mu U^\dagger D^\mu U \rangle^2
+L_2 \langle D_\mu U^\dagger D_\nu U \rangle 
     \langle D^\mu U^\dagger D^\nu U \rangle 
+L_3 \langle D^\mu U^\dagger D_\mu U D^\nu U^\dagger D_\nu U\rangle
\nonumber\\&&
+L_4 \langle D^\mu U^\dagger D_\mu U \rangle \langle \chi^\dagger U+\chi U^\dagger \rangle
+L_5 \langle D^\mu U^\dagger D_\mu U (\chi^\dagger U+U^\dagger \chi ) \rangle
+L_6 \langle \chi^\dagger U+\chi U^\dagger \rangle^2
\nonumber\\&&
+L_7 \langle \chi^\dagger U-\chi U^\dagger \rangle^2
+L_8 \langle \chi^\dagger U \chi^\dagger U + \chi U^\dagger \chi U^\dagger \rangle
-i L_9 \langle F^R_{\mu\nu} D^\mu U D^\nu U^\dagger +
               F^L_{\mu\nu} D^\mu U^\dagger D^\nu U \rangle
\nonumber\\&&
+L_{10} \langle U^\dagger  F^R_{\mu\nu} U F^{L\mu\nu} \rangle
+H_1 \langle F^R_{\mu\nu} F^{R\mu\nu} + F^L_{\mu\nu} F^{L\mu\nu} \rangle
+H_2 \langle \chi^\dagger \chi \rangle\,.
\ea
The constants $L_i$ in this Lagrangian are generally
known as Low-energy constants (LECs). The constants $H_i$
have values  dependent on the definition of currents/densities
and the terms are called contact terms. The LECs
absorb the divergences of loop diagrams 
order by order in the powercounting. The finite part is denoted
by $L_i^r$ and depends on the subtraction scale $\mu$ and the renormalization
prescription.

\emph{Number of parameters:} The number of parameters at the various orders
is shown in Tab.~\ref{tabparams}.
\begin{table}[t]
\begin{center}
\begin{tabular}{ccccccc}
      & \multicolumn{2}{c}{ 2 flavour} & \multicolumn{2}{c}{3 flavour} &
\multicolumn{2}{c}{ 3+3 PQChPT}\\
\hline
$p^2$ & $F,B$ & 2 & $F_0,B_0$ & 2 &  $F_0,B_0$ &  2 \\
$p^4$ & $l_i^r,h_i^r$ & 7+3 & $L_i^r,H_i^r$ & 10+2 & 
      $\hat L_i^r,\hat H_i^r$ &  11+2 \\
$p^6$ & $c_i^r$ & 52+4 & $C_i^r$ & 90+4 &  $K_i^r$ &
       112+3\\
\hline
\end{tabular}
\end{center}
\caption{\label{tabparams} The number of parameters+contact terms for the
various types of ChPT.}
\end{table}
The order $p^2$ is from \cite{Weinberg1}, order $p^4$ from \cite{GL0,GL1},
order $p^6$ from \cite{BCE1} after an earlier attempt \cite{FS2}.
The partially quenched results are derived from the $n_F$ flavour
case \cite{BDL1,BDL2}. The difficulty in obtaining a minimal set can be seen
from the recent discovery of a new relation in the $n_F=2$ case \cite{Haefeli1}.
Since the normal case is a continuous limit of the partially quenched
case, the resulting LECs are just linear combinations of
partially quenched LECs
using the Cayley-Hamilton relations given in \cite{BCE1}.
The general divergence structure at this order is also known \cite{BCE2}.
The parameters $B\ne B_0$ and $F\ne F_0$ are the two versus three-flavour
lowest order constants.

\emph{Chiral Logarithms:} The main predictions of ChPT are twofold. 1) It
relates processes with different numbers of pseudoscalars. 2) 
It predicts nonanalytic dependences at higher orders, often referred to
generically as \emph{Chiral Log(arithm)s}. As an example, the pion mass
for $n_F=2$ is given at NLO by \cite{GL0}
\be
\label{mpiNLO}
m_\pi^2 = 2 B \hat m  + \left(\frac{2 B \hat m}{F}\right)^2
\left[ \frac{1}{32\pi^2}{\log\frac{\left(2 B \hat m\right)}{\mu^2}} 
+ 2 l_3^r(\mu)\right] +\cdots
\ee
The notation $M^2 = 2 B \hat m\equiv B\left(m_u+m_d\right)$
is used a lot in the remainder.
The $\mu$ dependence cancels between the explicit dependence in the
logarithm and the implicit dependence in $l_3^r$.

\emph{LECs and choice of $\mu$:} The LECs, like $l_3^r$ in (\ref{mpiNLO}),
have to be determined experimentally or from lattice calculations.
They can be quoted in several ways. For $n_F=2$ Ref.~\cite{GL0} introduced
\be
\label{dellbar}
\bar l_i = \left(32\pi^2/\gamma_i\right)\, l_i^r(\mu)
-\log\left(M_\pi^2/\mu^2\right)\,,
\ee
which are $\mu$ independent and are proportional to the LECs $l_i^r(\mu=m_\pi)$.
For $n_F=3$ some of the corresponding $\gamma_i$ are zero and no good
equivalent definition of $\bar L_i$ exists. Here we always quote the
$L_i^r(\mu)$. The scale $\mu$ is in principle arbitrary but becomes relevant
when using estimates for higher order constants. Choosing
$\mu=m_\pi,m_K$ or $m_\eta$ puts some of the chiral logs to zero and thus
obscures one of the main predictions of ChPT. At a scale $\mu\approx 1$~GeV
experimentally $L_5^r\approx0$ and this would clash with large $N_c$ type
estimates of the LECs. For these reasons, many ChPT practitioners
use $\mu=m_\rho=0.77$~GeV.

\emph{What quantities to expand in:}
The ChPT expansion is in momenta and masses. However, one first has to decide
whether to expand in lowest order quantities, like $F,2B\hat m$, or
physical masses and decay constants, like $m_\pi,m_K,m_\eta,F_\pi,F_K$.
The latter is also not unique since relations like the
Gell-Mann--Okubo relation and kinematical relations like
$s+t+u=2m_\pi^2+2m_K^2$ for $\pi K$-scattering can be (and are heavily)
used to rewrite expressions. This sounds trivial but can change much how a
series convergence looks as shown below for a simple example. Similar
questions are also discussed by \cite{Descotes-Genon}. I personally prefer to
use physical masses and decay constants rather than the lowest
order quantities. The physical quantities are typically better known
and the chiral logs are created by particles propagating with their physical
momentum. Also, thresholds appear in the right places at each order in
perturbation theory.

\begin{figure}[ht]
\includegraphics[width=0.49\textwidth]{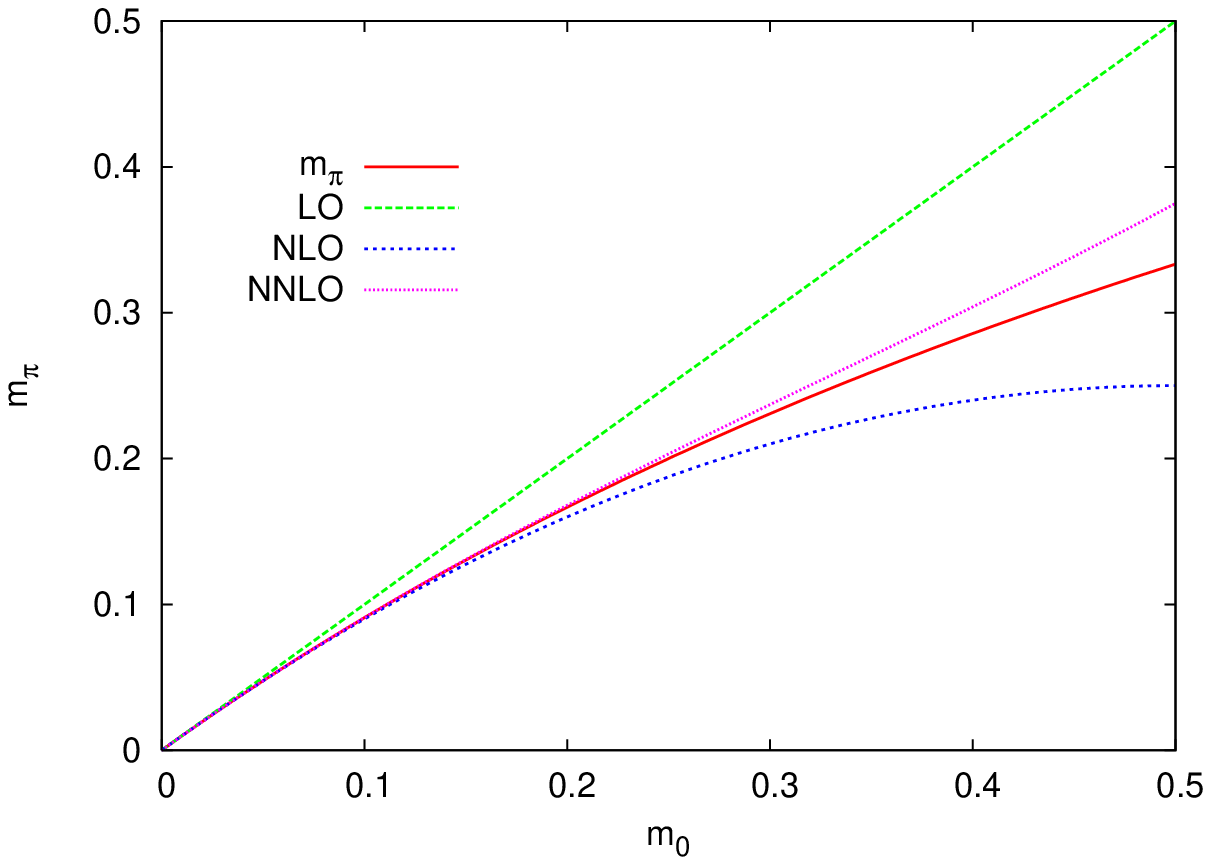}
\includegraphics[width=0.49\textwidth]{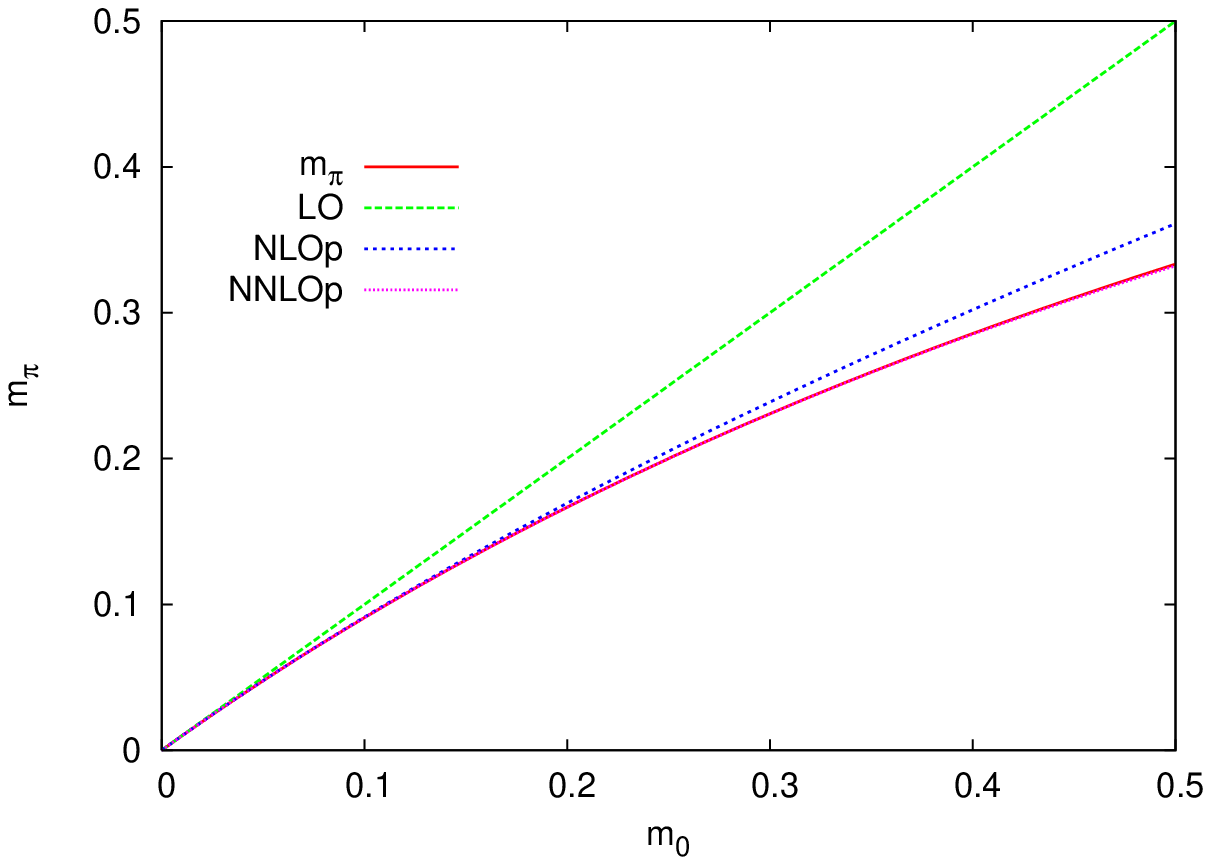}
\caption{On the left $m_\pi$ as a function of $m_0$ for the
expansion
in terms of $m_o/f_0$ 
(\protect\ref{mpi0}) 
and on the right for the expansion  
in terms
of $m_\pi/f_\pi$ 
(\protect\ref{mpi1}). Shown are the full results ($m_\pi$) and the first three
approximations.}
\label{figexample} 
\end{figure}

\emph{A very simple example:} Take the relations
$
m_\pi = m_0/\left(1+am_0/f_0\right),\quad
f_\pi = f_0/\left(1+bm_0/f_0\right)\,,
$
as exact. We can expand to NNLO in several ways
\ba
\label{mpi0}
m_\pi = m_0 -a \frac{m_0^2}{f_0} + a^2 \frac{m_0^3}{f_0^2}+\cdots
&\quad&
f_\pi = f_0 \left(1 -b \frac{m_0}{f_0} + b^2 \frac{m_0^2}{f_0^2}+\cdots\right)
\\
\label{mpi1}
 m_\pi = m_0 -a \frac{m_\pi^2}{f_\pi} + a (b-a) \frac{m_\pi^3}{f_\pi^2}+\cdots
&&
f_\pi = f_0 \left(1 -b \frac{m_\pi}{f_\pi} + b (2b-a) \frac{m_\pi^2}{f_\pi^2}+\cdots\right)
\\
\label{mpi2}
m_\pi = m_0 \left(1 -a \frac{m_\pi}{f_\pi} + a b \frac{m_\pi^2}{f_\pi^2}+\cdots\right)
\ea
The coefficients in the expansion and the actual numerical
values clearly depend on the way we write the results.
The plots in Fig.~\ref{figexample} show the convergence
for $a= 1$, $b= 0.5$ and $f_0=1$.
Only knowing the first three terms one would draw
very different conclusions on the quality of the convergence from
Fig.~\ref{figexample} for the different ways of writing the expansion.

\section{Two-flavour ChPT at NNLO}

References to order $p^2$ and $p^4$ work can be found in \cite{reviewp6}.
The first work at NNLO used dispersive methods to obtain the nonanalytic
dependence on kinematical quantities, $q^2,s,t,u$ at NNLO. This was done
for the vector (electromagnetic) and scalar formfactor of the pion
in \cite{GM} (numerically) and \cite{CFU} (analytically) and for
$\pi\pi$-scattering analytically in \cite{Knechtpipi}. The work of
\cite{Knechtpipi} allowed to put many of the full NNLO ChPT calculations
in two-flavour ChPT in a simple analytical form.

Essentially all processes of interest are calculated to NNLO fully in ChPT
starting with $\gamma\gamma\to\pi^0\pi^0$ \cite{BGS,GIS1},
$\gamma\gamma\to\pi^+\pi^-$ \cite{Burgi1,Burgi2,GIS2},
$F_\pi$ and $m_\pi$ \cite{Burgi2,BCEGS1,BCEGS2,BCT}, $\pi\pi$-scattering
\cite{BCEGS1,BCEGS2}, the pion scalar and vector formfactors \cite{BCT}
and pion radiative decay $\pi\to\ell\nu\gamma$ \cite{BT1}.
The pion mass is also known at order $p^6$ in finite volume \cite{CH}.

The LECs have been fitted in several processes. $\bar l_4$ from fitting
to the pion scalar radius \cite{BT1,CGL}, $\bar l_3$ from an estimate of
the pion mass dependence on the quark masses \cite{GL0,CGL}
and $\bar l_1$, $\bar l_2$ from the agreement with 
$\pi\pi$-scattering \cite{CGL}, $\bar l_6$ from the pion charge radius 
\cite{BCT}
and $\bar l_6-\bar l_5$ from the axial formfactor in $\pi\to\ell\nu\gamma$.
The final best values are \cite{BCT,BT1,CGL}
\be
\label{valueli}
\begin{array}{lll}
\bar l_1=-0.4\pm 0.6\,,\quad&
\bar l_2 =4.3\pm0.1\,,&
\bar l_3=2.9\pm2.4\,,
\\
\bar l_4=4.4\pm0.2\,,&
\bar l_6-\bar l_5 = 3.0\pm0.3\,,\quad&
\bar l_6 = 16.0\pm0.5\pm0.7\,.
\end{array}
\ee
It should be noticed that we do not have a good determination of $\bar l_3$
from the continuum.

There is also information on some combinations of $p^6$ LECs.
These are basically via the curvature in the vector and scalar formfactor
of the pion \cite{BCT} and two combinations from $\pi\pi$-scattering
\cite{CGL} from the knowledge of $b_5$ and $b_6$ in that reference.
The order $p^6$ LECs $c_i^r$ are estimated 
to have a small effect for
$m_\pi,f_\pi$ and $\pi\pi$-scattering.

\begin{figure}
\includegraphics[width=0.49\textwidth]{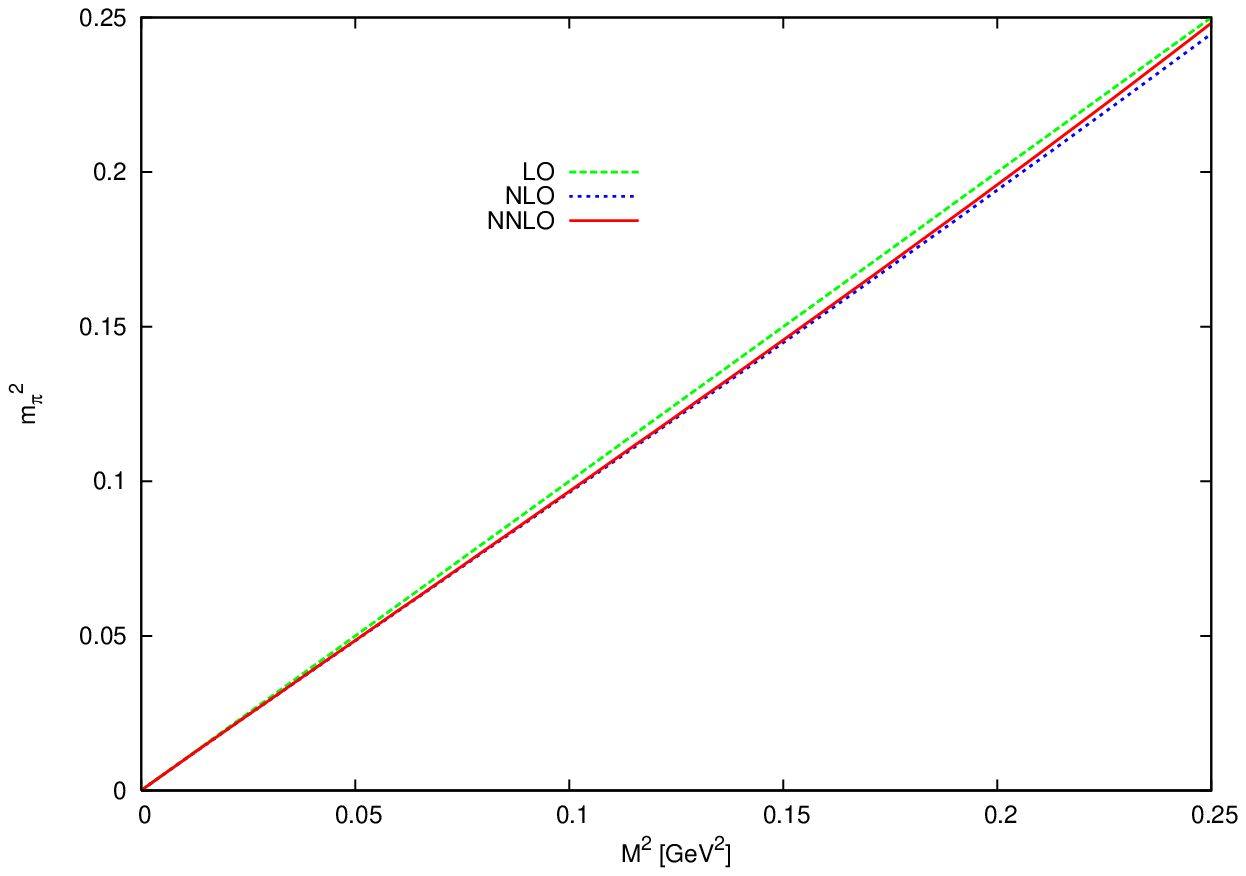}
\includegraphics[width=0.49\textwidth]{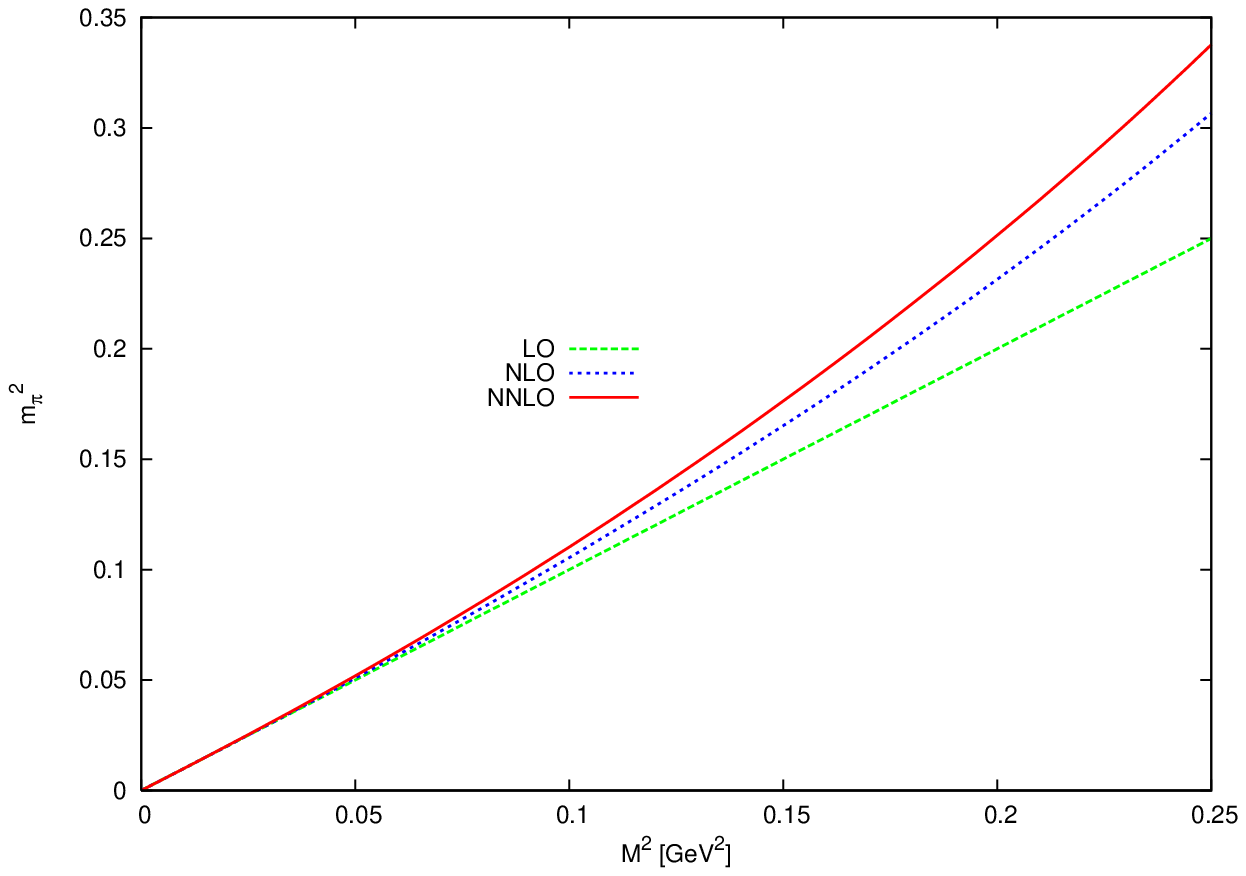}
\caption{The pion mass squared as a function of the quark mass via
$M^2=2B\hat m$, left with inputs as in (\protect\ref{valueli}) and right with
$\bar l_3=0$, both are for $n_F=2$ ChPT.}
\label{figmpi}
\end{figure}

\begin{figure}
\begin{center}
\includegraphics[width=0.49\textwidth]{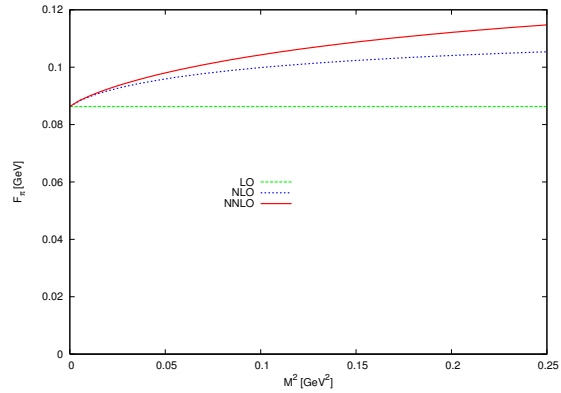}
\end{center}
\caption{The pion decay constant as a function of the quark mass via
$M^2=2B\hat m$, for $n_F=2$ ChPT.}
\label{figfpi}
\end{figure}

Let me now show a few dependences on the quark mass via $M^2 = 2B \hat m$.
First for $m_\pi^2$ expanded in analogy with (\ref{mpi2}).
A surprise is how small the NLO and NNLO corrections are for the values of
the input parameters given in (\ref{valueli}) and $c_i^r(\mu=0.77~\mathrm{GeV})=0$.
The full result is extremely linear as can be seen in the left plot in
Fig.~\ref{figmpi}
The linearity is a consequence of the fitting parameters as can be seen in
the right figure in Fig.~\ref{figmpi}. Similarly, $F_\pi$ as a function
of $M^2$ expanded as in (\ref{mpi1}) is shown in Fig.~\ref{figfpi}.
The values of $m_\pi^2$, $F_\pi$ and $M^2$ are determined selfconsistently
from the ChPT formulas quoted in \cite{BCT} via an iterative method.

\section{Three-flavour ChPT at NNLO}

\subsection{Calculations}

In this section I will discuss several results at NNLO in mesonic three-flavour
ChPT. In general the formulas here are much more involved than in
two-flavour ChPT and while the expressions have been reduced to a series
of well-defined two-loop integrals, the latter are evaluated numerically.
Most recent calculations use the subtraction scheme specified in
\cite{BCE2,BCEGS2} but many do not, also the reduction to numerical integrals
is done differently by different groups which makes comparisons very difficult
in the case of disagreement, see e.g. the discussion in \cite{BT3}
about the numerical discrepancy with \cite{PS3}.

The vector two-point functions were among the first calculated
in the $\pi$ and $\eta$ flavour quantum number channel \cite{ABT1,GK1}
and in the $K$  flavour quantum number channel \cite{ABT1,DK}.
The isospin breaking for the $\rho\omega$ channel was done in \cite{Maltman}.
The disconnected scalar two-point function
relevant for bounds on $L_4^r$ and $L_6^r$ was worked out in \cite{Moussallam}
The remaining scalar two-point functions are known but unpublished
\cite{Bijnensscalar}. Masses and decay constants as well as axial-vector
two-point functions were the first calculations which required full two-loop
integrals, done in the $\pi$ and $\eta$ \cite{ABT1,GK2} and the $K$ channel
\cite{ABT1}. The full isospin breaking contributions to masses and decay
constants are in \cite{ABT4}.

At this level many expressions were known but a full fit of LECs to
experimental data could be carried out only after $K_{\ell4}$ had also been
evaluated to NNLO \cite{ABT2,ABT3}. The vacuum expectation values
in the isospin limit were done in \cite{ABT3},
 with isospin breaking in \cite{ABT4} and at finite volume in~\cite{BG}.

The vector (electromagnetic) formfactors for pions and kaons were calculated
in \cite{PS1,PS2,BT2} and in \cite{BT2} a NNLO fit for $L_9^r$ was performed.
$L_{10}^r$ can be had from the axial formfactor in $\pi,K\to\ell\nu\gamma$.
The NNLO calculation is done, but no data fitting was performed\cite{Geng}.
A rather important calculation is the $K_{\ell3}$ formfactor. This calculation
was done by \cite{BT3,PS3} and a rather interesting relation between
the value at zero, the slope and the curvature for the scalar formfactor
obtained \cite{BT3}. I will present some results for $K_{\ell3}$ below.

Calculations for scalar formfactors including sigma terms and scalar radii
\cite{BD} and $\pi\pi$ \cite{BDT} and $\pi K$-scattering \cite{BDT2}
have been performed as well and used to place limits on $L_4^r$ and $L_6^r$.
Finally, the relations between the $l_i^r$ and $L_i^r$ has been extended
to the accuracy needed to compare order $p^6$ results in two and three-flavour
calculations \cite{Haefeli2} and there has been some progress towards fully
analytical results for
$m_\pi^2$ \cite{Kaiser} and $\pi K$-scattering lengths \cite{KS}.

I am aware of number of calculations in progress,
including $\eta\to3\pi$ and isospin breaking in $K_{\ell3}$ \cite{BG2},
preliminary results were reported in \cite{Kaon07},
the sunsetintegrals needed for the masses at finite volume
\cite{BL3} and the relations
between the two and three flavour order $p^6$ constants \cite{Haefeli3}.

\subsection{$C_i^r$: estimates of order $p^6$ LECs}

\begin{figure}[ht]
\begin{center}
\unitlength=0.5pt
\begin{picture}(440,100)
\SetScale{0.5}
\SetWidth{1.5}
\Line(0,100)(10,50)
\Line(0,0)(10,50)
\Text(10,10)[]{$\pi$}
\Text(10,90)[]{$\pi$}
\Vertex(10,50){5}
\Line(10,52)(100,52)
\Line(10,48)(100,48)
\Text(55,65)[]{$\rho,S$}
\Text(55,30)[]{\small$\rightarrow q^2$}
\Vertex(100,50){5}
\Line(100,50)(110,0)
\Line(100,50)(110,100)
\Text(120,10)[]{$\pi$}
\Text(120,90)[]{$\pi$}
\Text(220,75)[]{\tiny$|q^2|<< m_\rho^2,m_S^2$}
\Text(220,50)[]{\Large$\Longrightarrow$}
\Line(320,100)(380,50)
\Line(320,0)(380,50)
\Line(380,50)(440,100)
\Line(380,50)(440,0)
\Text(380,75)[]{\tiny$C_i^r$}
\Vertex(380,50){5}
\end{picture}
\end{center}
\caption{A schematic indication of the estimate of the order $p^6$ LECs by resonance exchange.}
\label{figCi}
\end{figure}
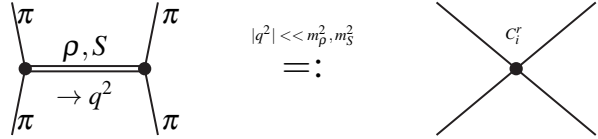

Most numerical analysis at order $p^6$ use a (single) resonance
approximation to the order $p^6$ LECs. This is schematically shown in
Fig.~\ref{figCi}.
The main underlying motivation is the large $N_c$ limit and phenomenological
success at order $p^4$ \cite{Ecker1}.
There is a large volume of work on this,
some references are \cite{MHA1,MHA2,BGLP,Reso4}.
The numerical work I will report has used the simple resonance Lagrangian
\cite{BCEGS2,ABT4,ABT3,Ecker1}
\ba
\label{lagreso}
{\cal L}_V &=& -\frac{1}{4}\langle V_{\mu\nu}V^{\mu\nu}\rangle
+\frac{1}{2}m_V^2\langle V_\mu V^\mu\rangle
-\frac{f_V}{2\sqrt{2}}\langle V_{\mu\nu}f_+^{\mu\nu}\rangle
-\frac{ig_V}{2\sqrt{2}}\langle V_{\mu\nu}[u^\mu,u^\nu]\rangle
+f_\chi\langle V_\mu[u^\mu,\chi_-]\rangle\,,
\nonumber\\
{\cal L}_A &=& -\frac{1}{4}\langle A_{\mu\nu}A^{\mu\nu}\rangle
+\frac{1}{2}m_A^2\langle A_\mu A^\mu\rangle
-\frac{f_A}{2\sqrt{2}}\langle A_{\mu\nu}f_-^{\mu\nu}\rangle\,,
\nonumber\\
{\cal L}_S &=& \frac{1}{2} \langle \nabla^\mu S \nabla_\mu S 
 - M^2_S S^2 \rangle  
 + c_d \langle Su^\mu u_\mu \rangle + c_m \langle S \chi_+ \rangle\,, 
\,\,\,
{\cal L}_{\eta^\prime} = \frac{1}{2}\partial_\mu P_1\partial^\mu P_1
-\frac{1}{2}M_{\eta^\prime}^2 P_1^2
+i\tilde d_m P_1\langle\chi_- \rangle\,.
\nonumber\\
f_V &=& 0.20,\quad  f_\chi = -0.025,\quad  g_V = 0.09, 
\quad  c_m = 42 \mbox{ MeV},\quad
  c_d = 32 \mbox{ MeV}, \quad
 \tilde d_m = 20 \mbox{ MeV}, 
\nonumber\\
m_V &=& m_\rho = 0.77 \mbox{ GeV},  m_A = m_{a_1} = 1.23 \mbox{ GeV}, 
m_S = 0.98 \mbox{ GeV},\quad m_{P_1} =  0.958 \mbox{ GeV}\,.
\ea
The values of $f_V$, $g_V$, $f_\chi$ and $f_A$ come from experiment
 \cite{BCEGS2,Ecker1} and
$c_m$ and $c_d$ from resonance saturation at order $p^4$ \cite{Ecker1}. 

The estimates of the $C_i^r$ is the weakest point in the numerical
fitting at present, however, many results are not very sensitive to this.
The main problem is how the $C_i^r$ which contributes to the masses,
estimated to be zero except for $\eta'$ effects by (\ref{lagreso}),
affect the determination of the others. The estimate is also $\mu$-independent
while the $C_i^r$ depend on $\mu$.

The fits done here in \cite{ABT4,ABT2,ABT3,BD} try to check this by varying
the total resonance contribution by a factor of two, varying the scale $\mu$
from $550$ to $1000$~MeV and compare estimated $C_i^r$ to experimentally
determined ones. The latter works well, but again the experimentally 
well determined ones are those with dependence on kinematic variables only,
not ones relevant for quark-mass dependence.

\subsection{The fitting and results}

The inputs used for the fitting, see the more extensive discussion in
\cite{ABT4,ABT3}, are
\begin{itemize}
\parskip0cm\itemsep0cm
\item $K_{\ell4}$: $F(0)$, $G(0)$, $\lambda$ from E865 at 
BNL\cite{Pislak1,Pislak2}.
\item $m^2_{\pi^0}$, $m^2_\eta$, $m_{K^+}^2$, $m_{K^0}^2$, electromagnetic
corrections include the estimated violation of Dashen's theorem
(\cite{Dashen} and references therein). 
\item $F_{\pi^+}$.
\item $F_{K^+}/F_{\pi^+}$.
\item
$m_s/\hat m = 24$. Variations with
$m_s/\hat m$ were studied in \cite{ABT4,ABT3}.
\item
$L_4^r, L_6^r$ the main fit, 10, has them equal to zero, but see below
and the arguments in \cite{Moussallam}.
\end{itemize}
The results of this fit are summarized in Tab.~\ref{tabfits}.
The errors are very correlated, this is shown in Fig.~6 in \cite{ABT3} for the
fit to then available $K_{\ell4}$ data. As said before varying the estimates
of the $C_i^r$ by a factor of two, varying the $\mu$ where the estimate
of the $C_i^r$ is done from 550 to 1000~MeV all stay within the given
errors. Varying the values of $L_4^r,L_6^r$ as input can be done with a
reasonable fitting chi-squared when varying $10^3 L_4^r$ from $-0.4$ to $0.6$
and $L_6^r$ from $-0.3$ to $0.6$. These alternative fits were performed in
\cite{BD} and variation of many quantities with $L_4^r,L_6^r$
 (including the changes via the changed values of the other $L_i^r$) are shown
in \cite{BD,BDT,BDT2}. Fit B was one of the fits that gave a good fit to the
pion scalar radius and fairly small corrections to the sigma terms \cite{BD}
while fit D \cite{Kazimierz} is the one that gave agreement with
$\pi\pi$ and $\pi K$-scattering threshold quantities.

One point should be observed here, if one fits lattice data with NLO formulas
to obtain the $L_i^r$, one should also use the NLO or order $p^4$ fit values
from Tab.~\ref{tabfits} to compare with. 

Note the $m_u/m_d=0$ is never even
close to the best fit and this remains true for the entire variation
with $L_4^r,L_6^r$. The value of $F_0$,
the pion decay constant in the three-flavour chiral limit,
can vary significantly, even though I believe that fit B is an extreme case.

\begin{figure}[htb]
\vskip-0.5cm
\includegraphics[angle=-90,width=0.49\textwidth]{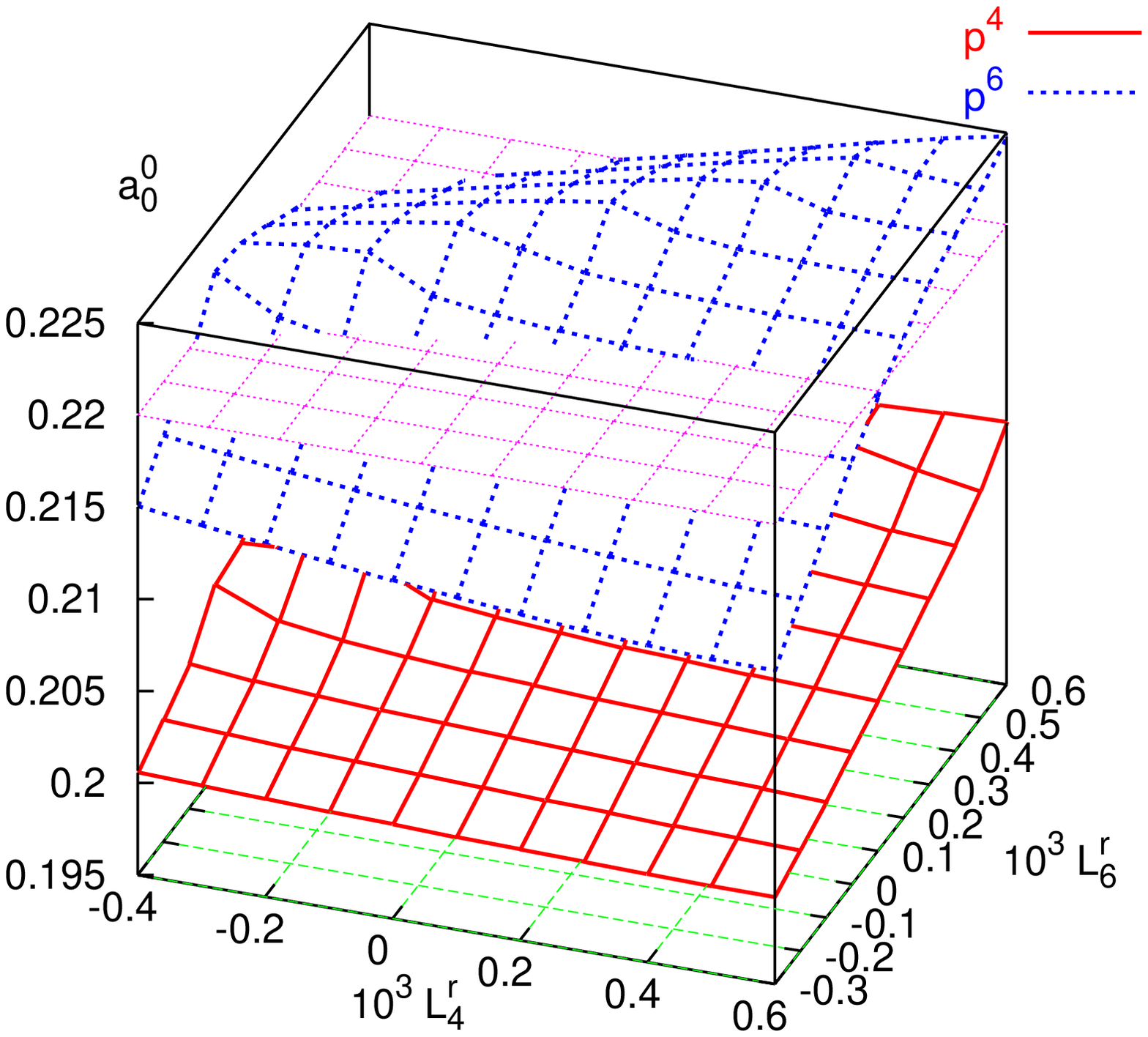}
\includegraphics[angle=-90,width=0.49\textwidth]{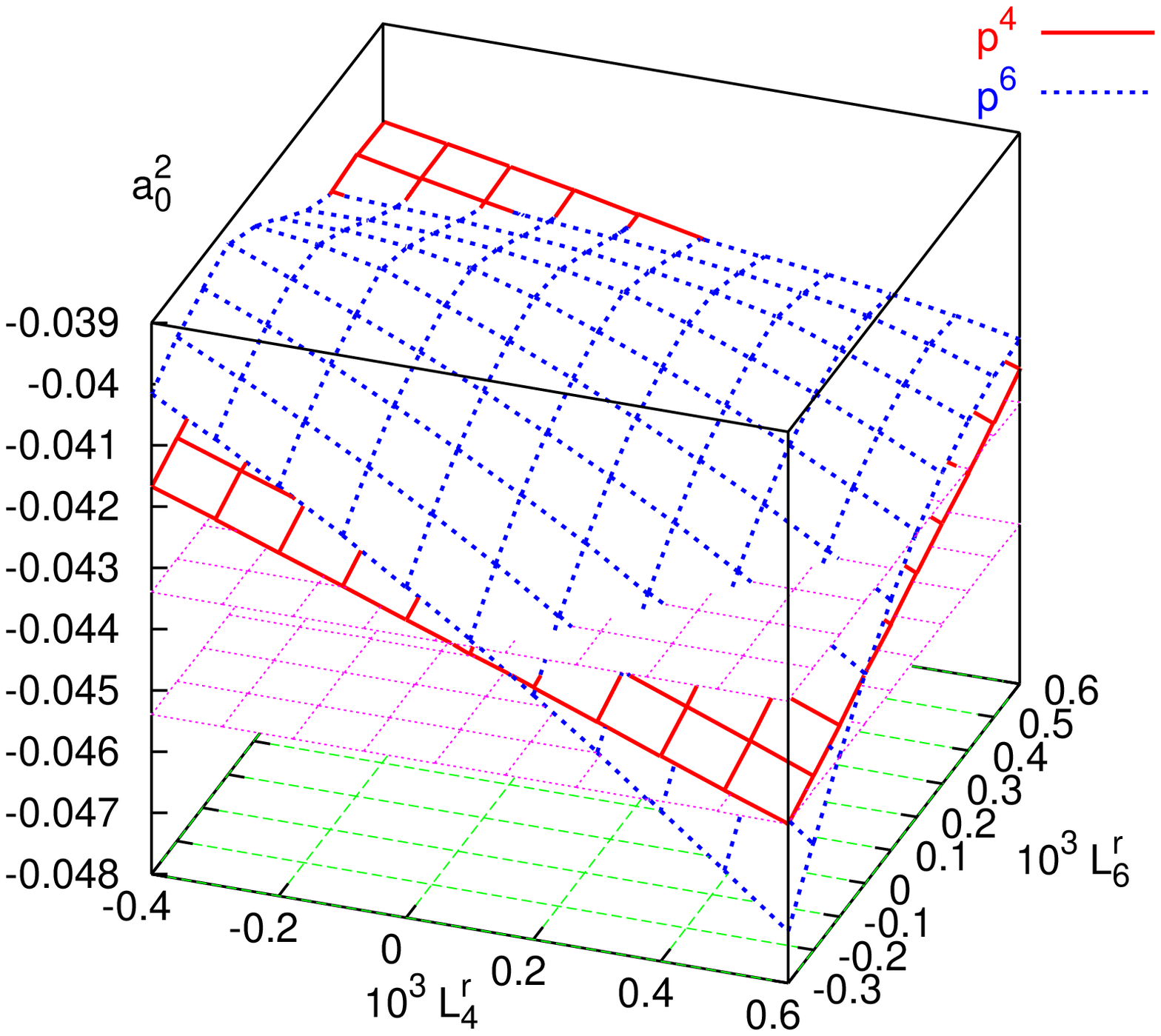}
\caption{The $\pi\pi$ scattering lengths in three-flavour ChPT as
a function of the input values of $L_4^r,L_6^r$ used in the fits. 
$a^0_0$ left, $a^2_0$ right.
See \cite{BDT}
for details, from \cite{BDT}.}
\label{figpipi}
\end{figure}

\begin{figure}[ht]
\begin{center}
\unitlength=0.4pt
\begin{overpic}[angle=-90,origin=rb,width=0.4\textwidth]{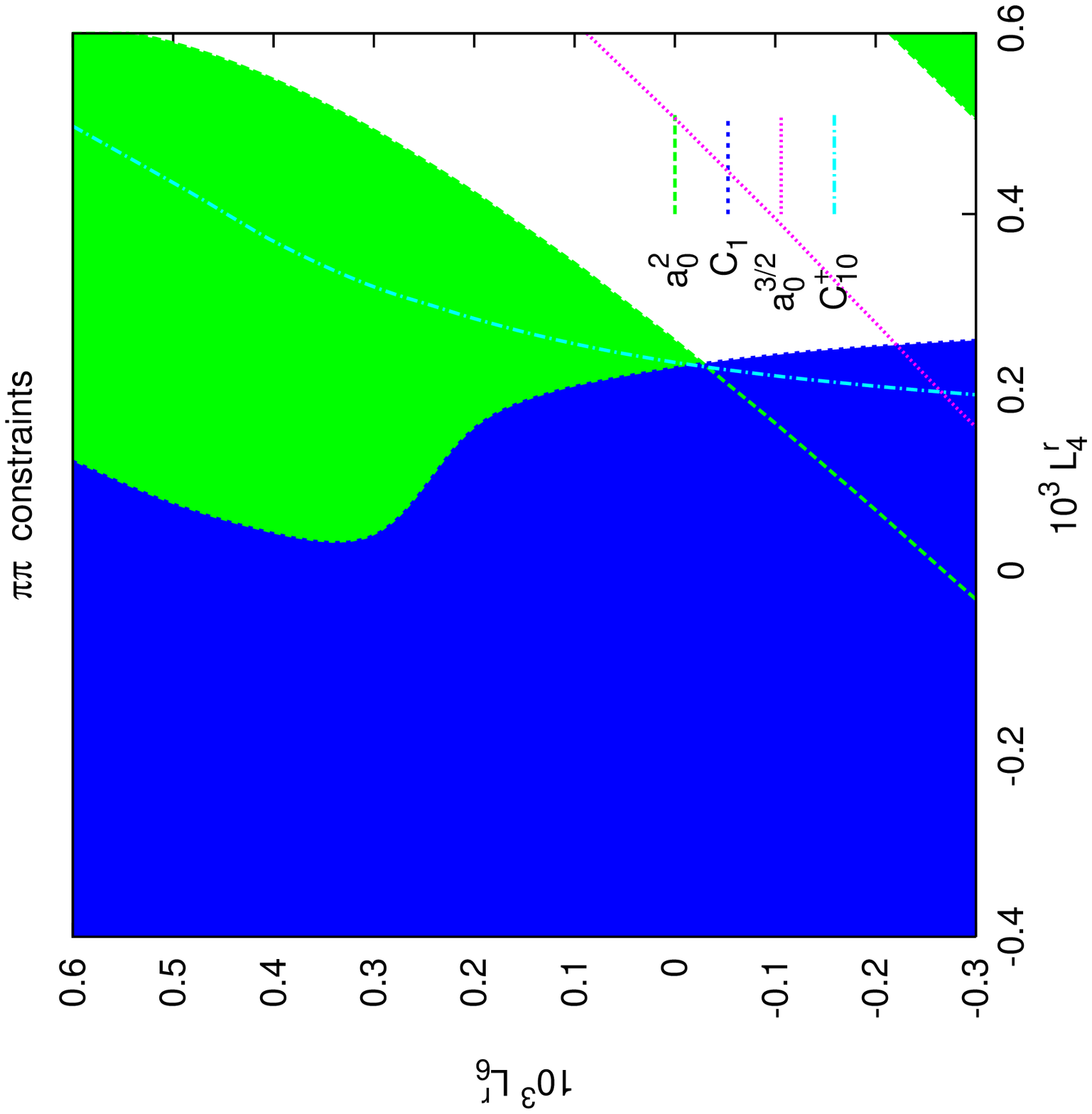}
\SetScale{0.4}
\SetWidth{1.5}
\CArc(285,155)(20,0,360)
\end{overpic}
\unitlength=0.4pt
\begin{overpic}[angle=-90,width=0.4\textwidth]{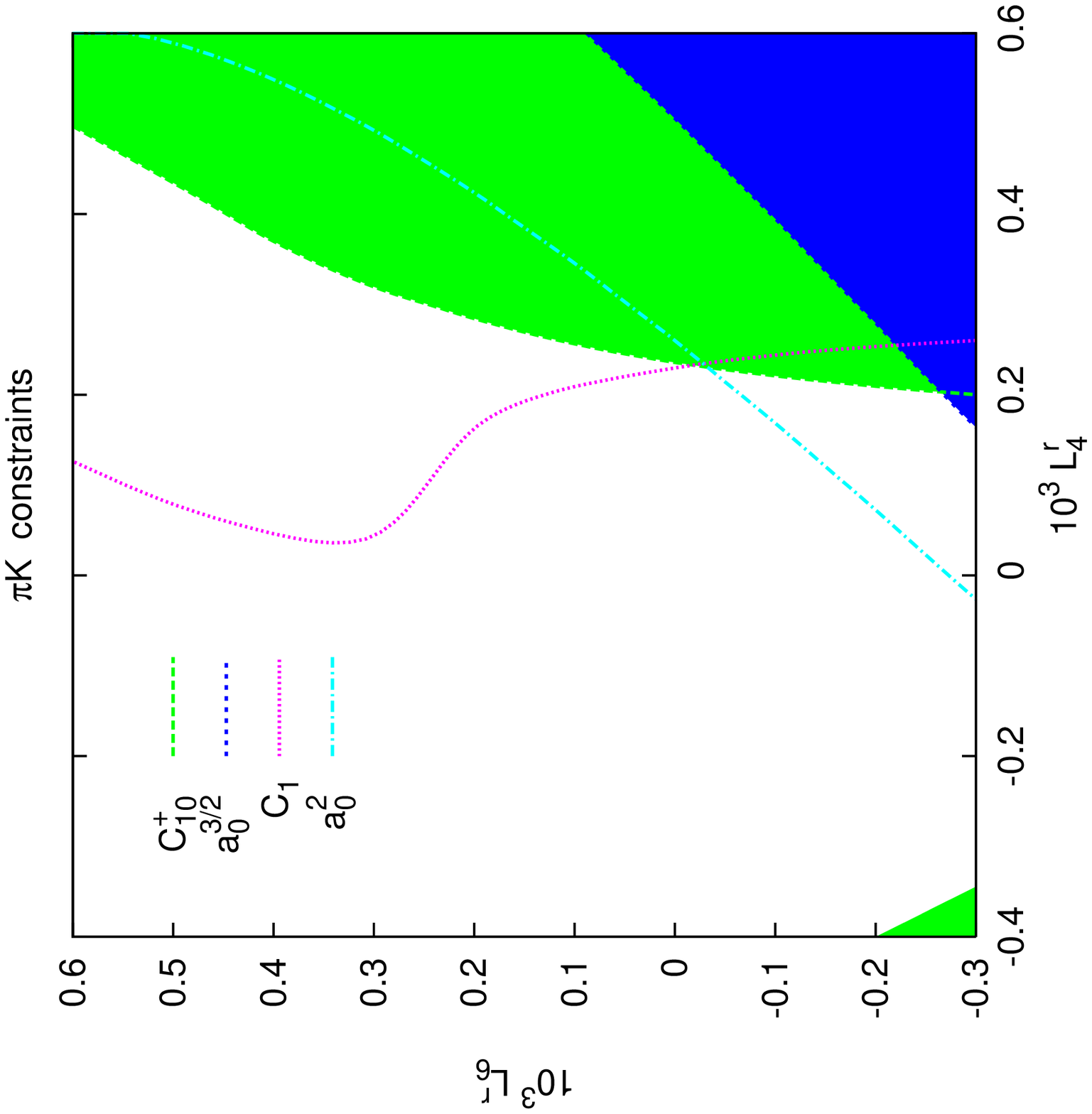}
\SetScale{0.4}
\SetWidth{1.5}
\CArc(275,160)(20,0,360)
\end{overpic}
\end{center}
\caption{The bounds on $L_4^r,L_6^r$ from $\pi\pi$ and $\pi K$-scattering
threshold parameters. Left $\pi\pi$ where the bound from $a^2_0$
shown in Fig.~\protect\ref{figpipi} is the most stringent.
Right $\pi K$. 
White regions are allowed.
The region of fit D, compatible with both, is indicated
by the circle. From \protect\cite{BDT2}.}
\label{figL4L6}
\end{figure}

In Fig.~\ref{figpipi} we show how the threshold parameters $a_0^0$ and $a^0_0$
depend on the variation with  $L_4^r,L_6^r$. $a_0^0$ always agrees
well with the result of \cite{CGL} while $a^2_0$ only agrees well within a
limited region \cite{BDT}. For comparison, the order $p^2$ values are
$a_0^0=0.159$ and $a^2_0=-0.0454$. The planes in Fig.~\ref{figpipi} indicate
the results $a_0^0=0.220\pm0.005$, $a^2_0= -0.0444\pm0.0010$ \cite{CGL}.
\begin{table}[ht]
\begin{center}
\begin{tabular}{ccccc}
                & fit 10 & same $p^4$ & fit B & fit D\\
\hline
$10^3 L_1^r$ & $0.43\pm0.12$ & $0.38$ & $0.44$ & $0.44$\\
$10^3 L_2^r$ & $0.73\pm0.12$ & $1.59$ & $0.60$ & $0.69$\\
$10^3 L_3^r$ & $-2.53\pm0.37$ & $-2.91$ &$-2.31$&$-2.33$\\
$10^3 L_4^r$ & $\equiv0$    & $\equiv 0$& $\equiv0.5$ & $\equiv0.2$\\
$10^3 L_5^r$ & $0.97\pm0.11$& $1.46$ & $0.82$ & $0.88$\\
$10^3 L_6^r$ & $\equiv0$    & $\equiv 0$& $\equiv0.1$ & $\equiv0$\\
$10^3 L_7^r$ & $-0.31\pm0.14$&$-0.49$ & $-0.26$ & $-0.28$\\
$10^3 L_8^r$ & $0.60\pm0.18$ & $1.00$ & $0.50$ & $0.54$\\
$10^3 L_9^r$ & $5.93\pm0.43$ & $7.0$  & --      &  -- \\
\hline
$2 B_0 \hat m/m_\pi^2$ & 0.736 & 0.991 & 1.129 & 0.958\\
$m_\pi^2$: $p^4,p^6$    & 0.006,0.258 & 0.009,$\equiv0$ & $-$0.138,0.009 &
          $-$0.091,0.133\\
$m_K^2$: $p^4,p^6$    & 0.007,0.306 & 0.075,$\equiv0$ & $-$0.149,0.094 &
          $-$0.096,0.201\\
$m_\eta^2$: $p^4,p^6$    & $-$0.052,0.318 & 0.013,$\equiv0$ & $-$0.197,0.073 &
          $-$0.151,0.197\\
$m_u/m_d$    & 0.45$\pm$0.05 & 0.52 & 0.52 & 0.50\\
\hline
$F_0$ [MeV]          & 87.7 & 81.1 & 70.4 & 80.4 \\
$\frac{F_K}{F_\pi}$: $p^4,p^6$ & 0.169,0.051 & 0.22,$\equiv0$ & 0.153,0.067 &
   0.159,0.061
\end{tabular}
\end{center}
\caption{The fits of the $L_i^r$ and some results, see text for
a detailed description. They are all quoted at $\mu=0.77$~GeV.
Table with values from \protect\cite{ABT4,BT2,BD,BDT2,Kazimierz}.
At present the best fits to use for comparison with the lattice
are fit 10 at NNLO or order $p^4$ depending on the order of the lattice fit.}
\label{tabfits}
\end{table}
The same study was performed for $\pi K$ scattering lengths in \cite{BDT2}
with the results of the Roy-Steiner analysis \cite{BDM}. The resulting
limits on the input values of  $L_4^r,L_6^r$ are shown in Fig.~\ref{figL4L6}.
The resulting region called fit D in Tab.~\ref{tabfits} is
 $10^3 L_4^r \approx0.2$, $10^3 L_6^r \approx0.0$.
This general fitting obviously needs more work and systematic studies
and constraints from lattice QCD on $L_4^r,L_6^r$ will be very useful.

\subsection{Mass dependence and other results for selected quantities}

I now show the dependence of a few quantities on the input masses.
These are updates of the plots shown in \cite{ABT3} and also some new
ones for $f_+(0)$ in $K_{\ell3}$. The masses squared and decay constants
are written in the form analogous to (\ref{mpi1}) as published
in \cite{ABT1} (note the erratum of \cite{ABT4} and the formulas
given in \cite{website}).
A selfconsistent set of $m_\pi^2$, $m_K^2$, $m_\eta^2$, $F_\pi$, $B_0 m_s$
and $B_0 \hat m$ with the fitted values of $L_i^r$ and $F_0$ is determined for
each input value of two masses. This is done by iterating the formulas
till convergence is reached. NNLO reproduces the physical values
at the physical point.

\begin{figure}
\includegraphics[width=0.49\textwidth]{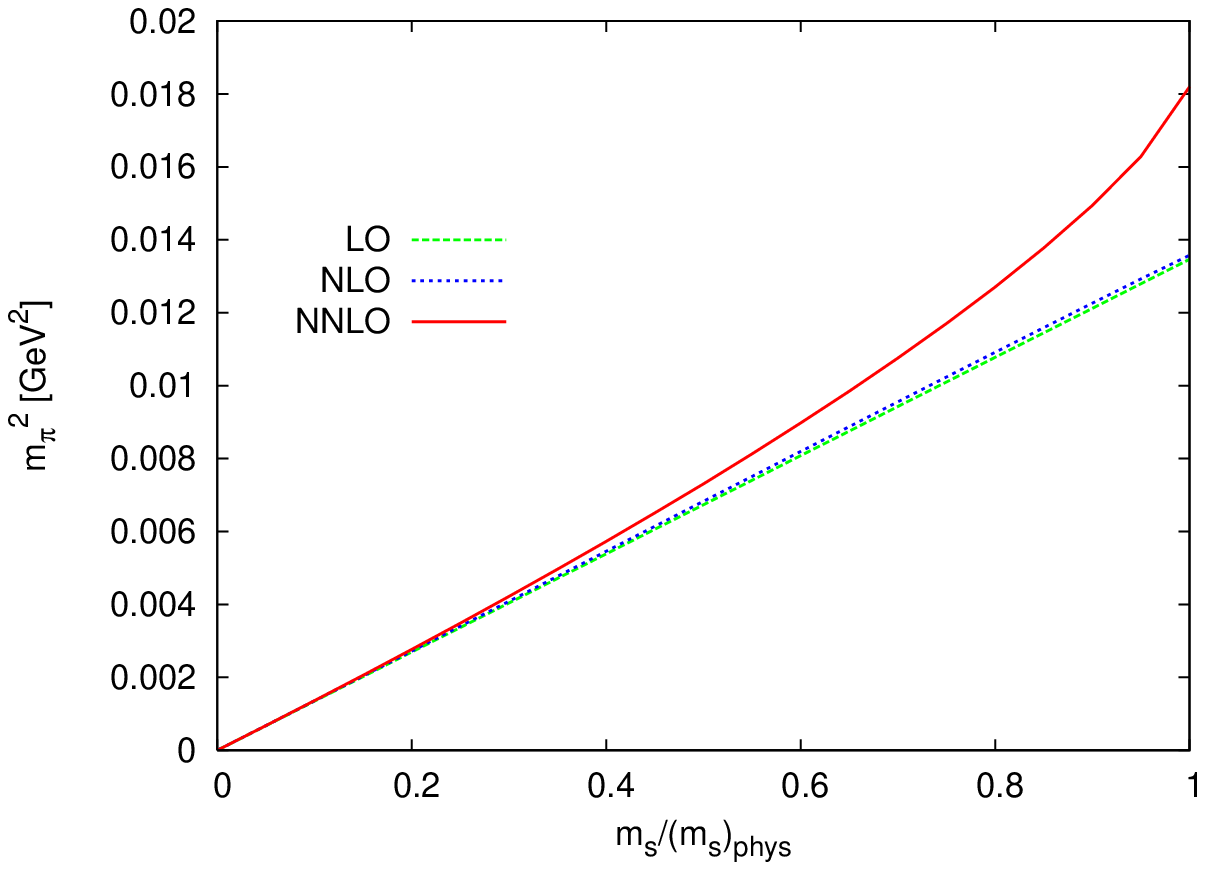}
\includegraphics[width=0.49\textwidth]{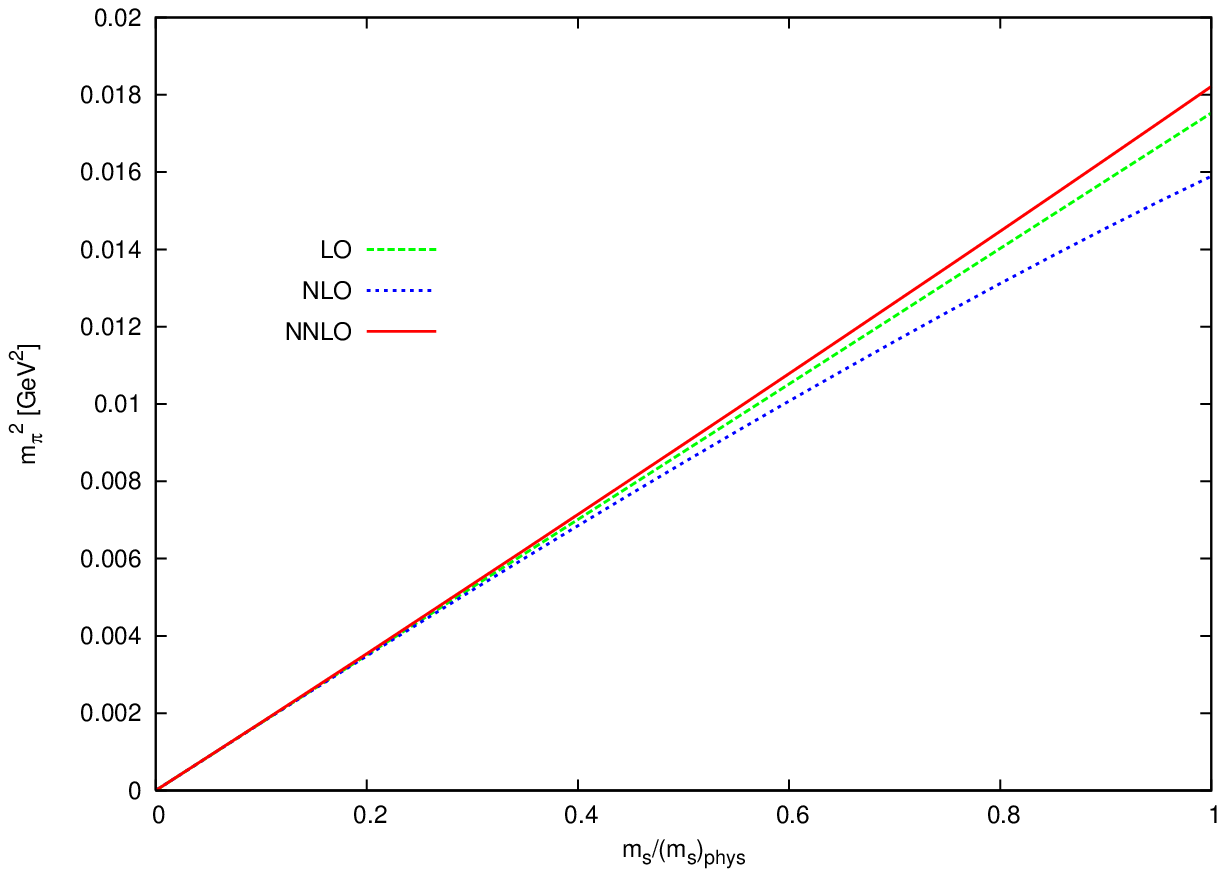}
\caption{$m_\pi^2$ as a function of $m_s$ for fit 10 
(left) and fit D (right) of 
Tab.~\protect\ref{tabfits} with $m_s/\hat m$ fixed. 
Note the difference in convergence properties between the two fits.}
\label{figmpinf3}
\end{figure}

\begin{figure}
\includegraphics[width=0.49\textwidth]{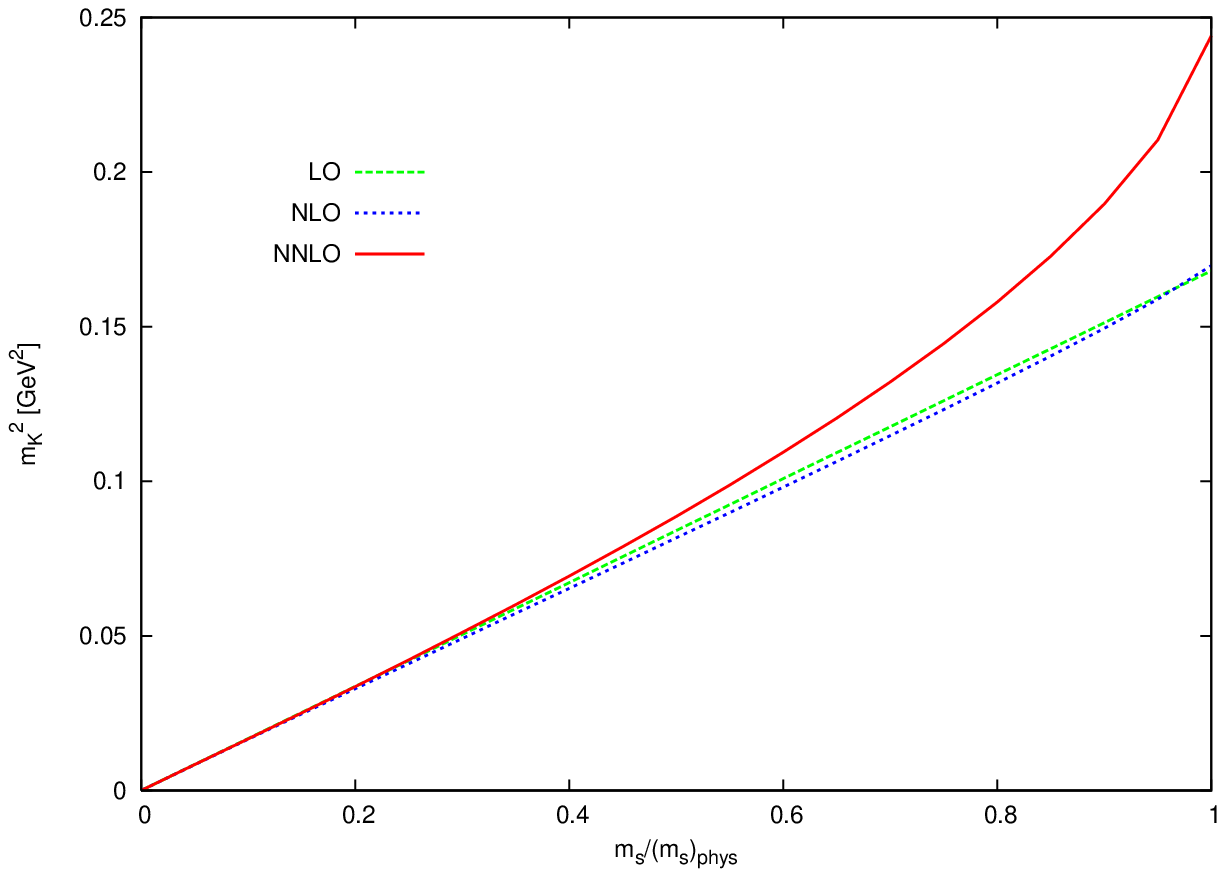}
\includegraphics[width=0.49\textwidth]{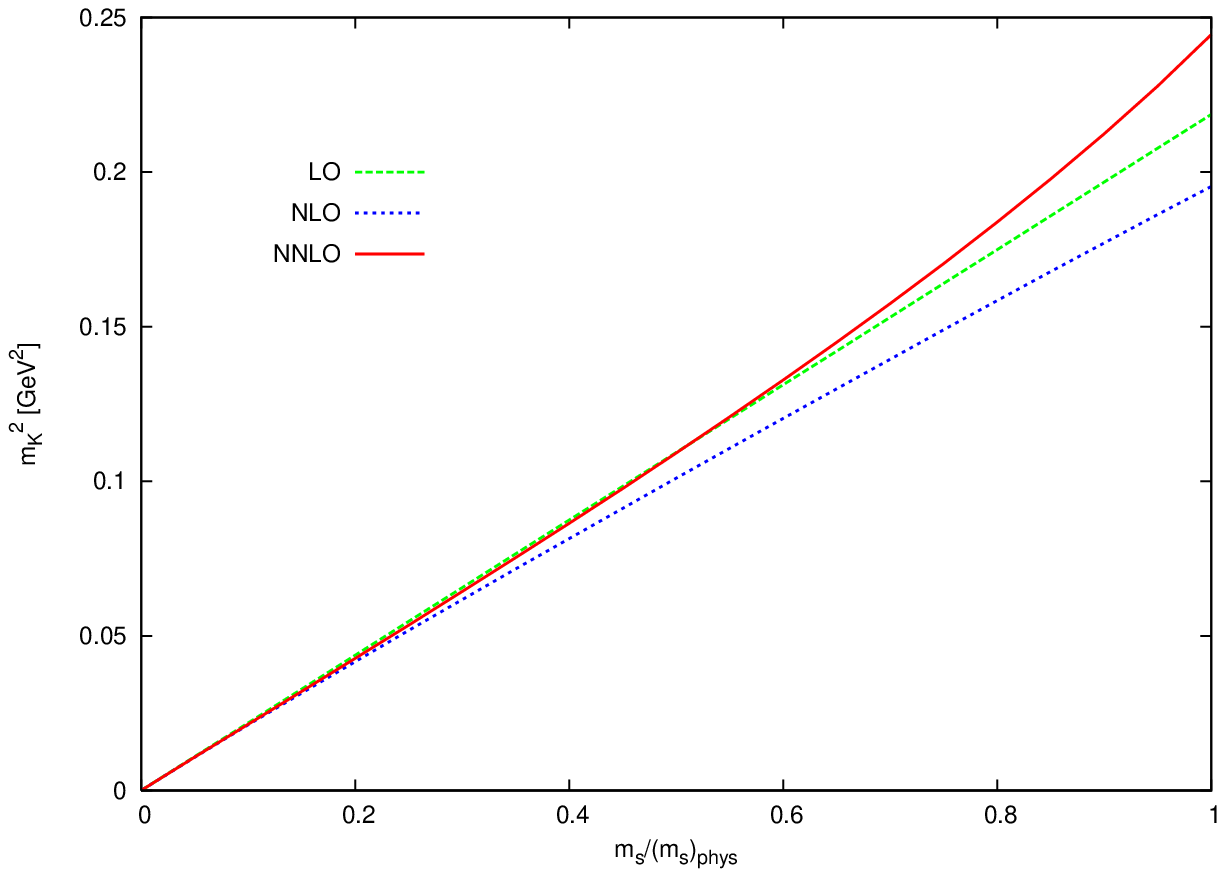}
\caption{$m_K^2$ as a function of $m_s$ for fit 10 (left) and fit D (right) 
of Tab.~\protect\ref{tabfits} with $m_s/\hat m$ fixed. 
Note the difference in convergence properties between the two fits.}
\label{figmknf3}
\end{figure}

\begin{figure}
\begin{minipage}{0.49\textwidth}
\includegraphics[width=\textwidth]{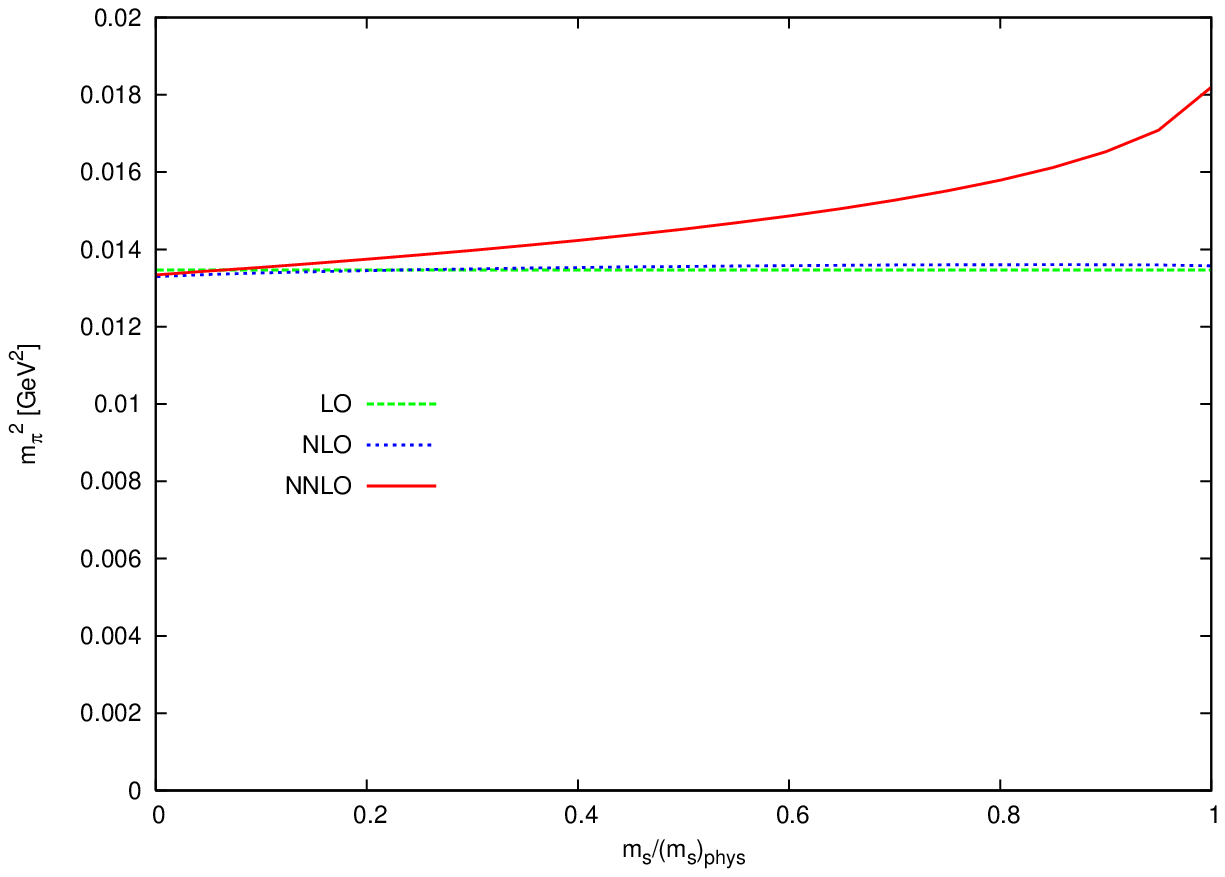}
\centerline{(a)}
\end{minipage}
\begin{minipage}{0.49\textwidth}
\includegraphics[width=\textwidth]{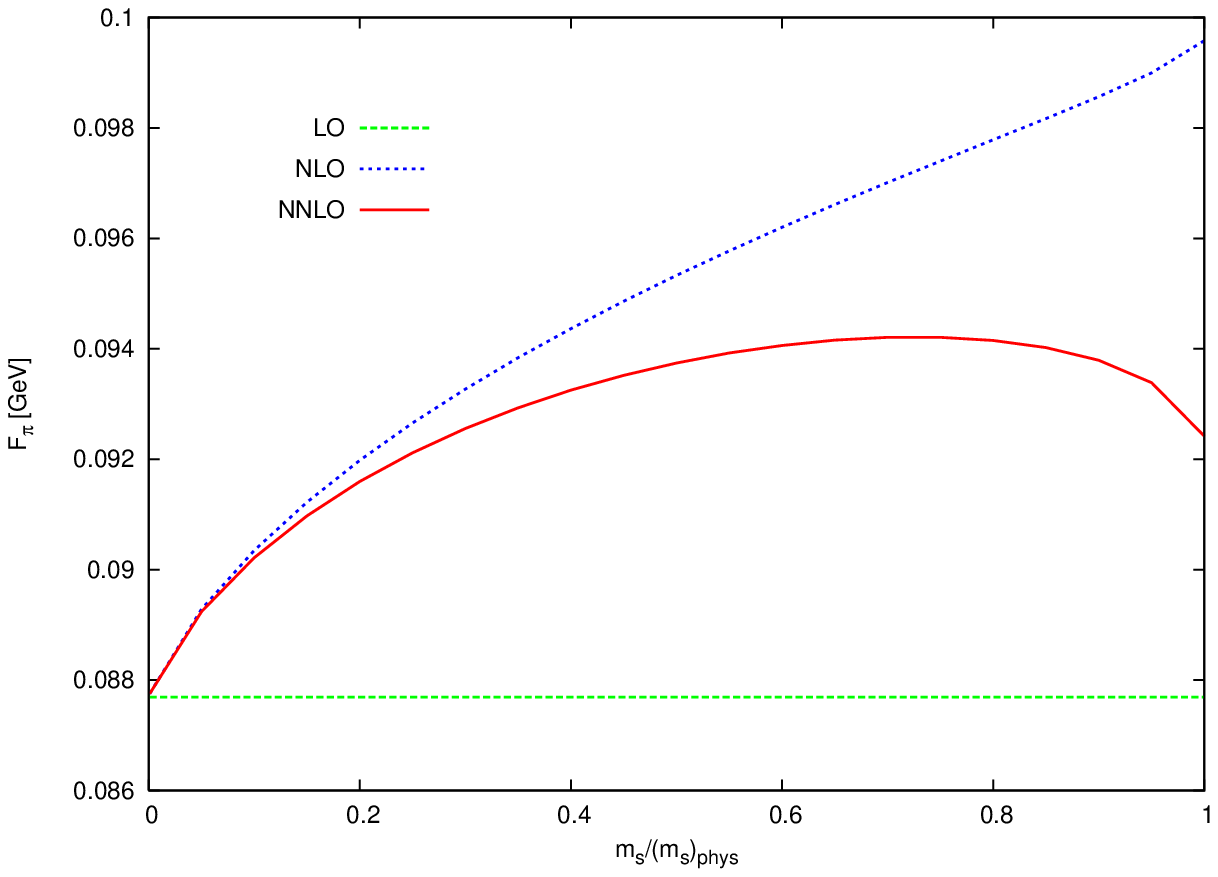}
\centerline{(b)}
\end{minipage}
\caption{For fit 10: (a) $m_\pi^2$ as a function of $m_s$ with $\hat m$ fixed.
(b) $F_\pi$ as a function of $m_s$ with $m_s/\hat m$ fixed.}
\label{figmpimq}
\end{figure}

We show $m_\pi^2$ for fit 10 and fit D keeping $m_s/\hat m=24$ and varying
$m_s$ in Fig.~\ref{figmpinf3}. The same dependence but for $m_K^2$ is shown in Fig.~\ref{figmknf3}. The large corrections for fit 10 come from the kaon mass.
This is shown in Fig.~\ref{figmpimq}(a) where we plot $m_\pi^2$
with $\hat m$ fixed to its physical value and vary $m_s$.

\begin{figure}
\begin{minipage}{0.49\textwidth}
\includegraphics[width=\textwidth]{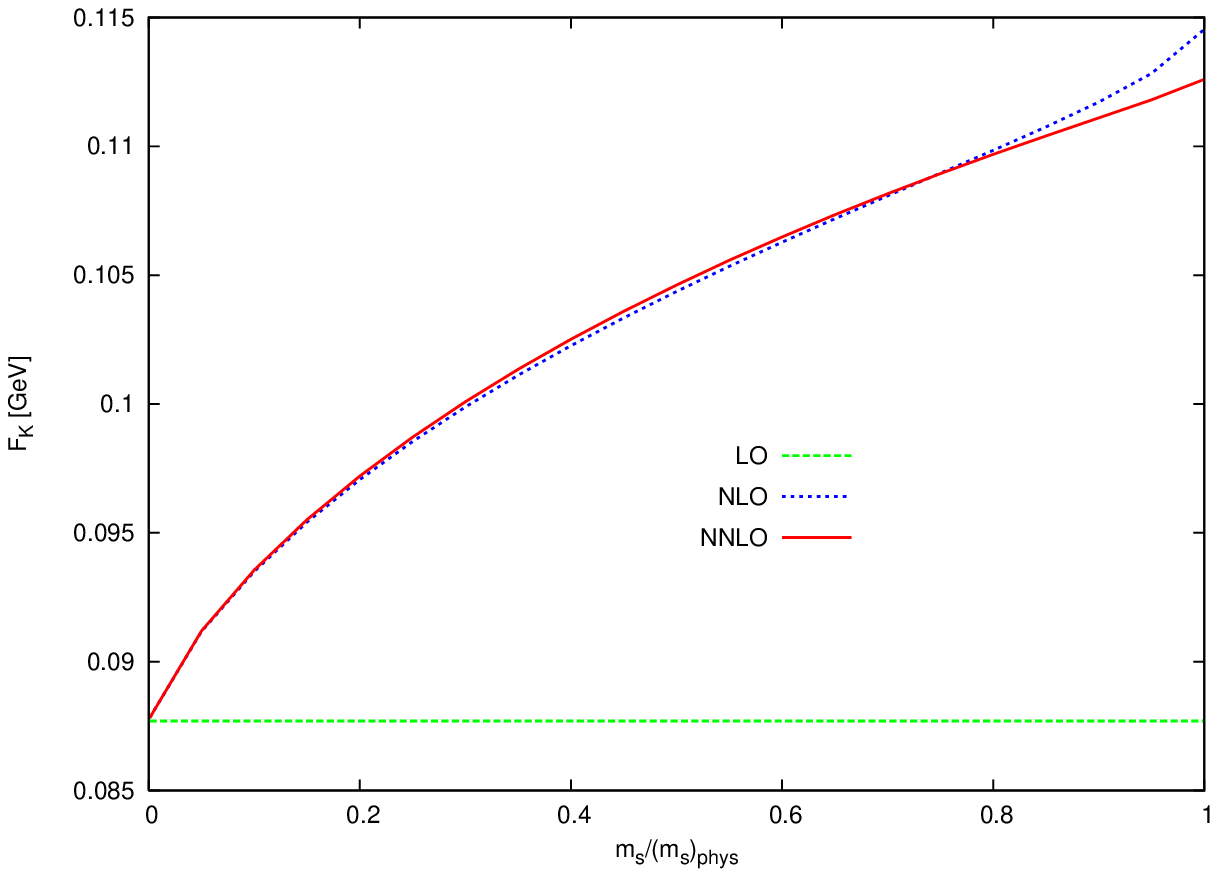}
\centerline{(a)}
\end{minipage}
\begin{minipage}{0.49\textwidth}
\includegraphics[width=\textwidth]{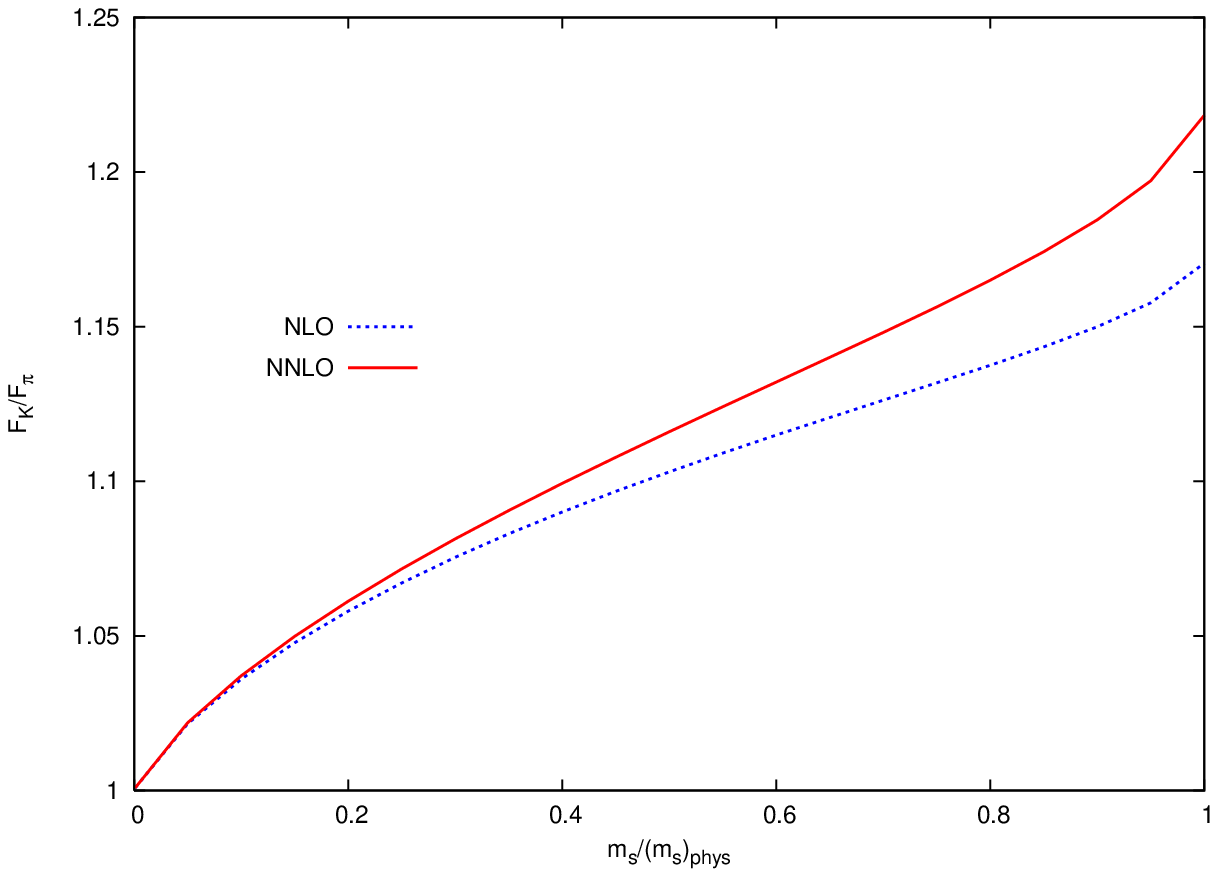}
\centerline{(b)}
\end{minipage}
\caption{For fit 10: (a) $F_K$ (b) $F_K/F_\pi$ as a function of $m_s$
 with $m_s/\hat m$ fixed.}
\label{figfkmq}
\end{figure}

The decay constants and ratios are shown as a function of $m_s$ at fixed ratio
$m_s/\hat m$ for $F_\pi$ in Fig.~\ref{figmpimq}(b), 
$F_K$ in Fig.~\ref{figfkmq}(a) and $F_K/F_\pi$ in  Fig.~\ref{figfkmq}(b)

Let me finally discuss some results of the calculation of $K_{\ell3}$ \cite{BT3}.
One major point is that the scalar formfactor can be written as
\ba
f_0(t) &=& 1-\frac{8}{F_\pi^4}{\left(C_{12}^r+C_{34}^r\right)}
\left(m_K^2-m_\pi^2\right)^2
+\frac{8t}{F_\pi^4}{\left(2C_{12}^r+C_{34}^r\right)}
\left(m_K^2+m_\pi^2\right)
-\frac{8}{F_\pi^4} t^2 {C_{12}^r}
+\tilde \Delta(t)\,.
\ea
$\tilde\Delta(t)$ contains \emph{no} $C_i^r$ and only depends on the
$L_i^r$ at order $p^6$.
\emph{All needed parameters $C_i^r$ for $f_+(0)=f_0(0)$
can thus be determined experimentally or from the lattice} via the slope
and curvature of $f_0(t)$.

\begin{figure}
\begin{minipage}{0.49\textwidth}
\includegraphics[width=\textwidth]{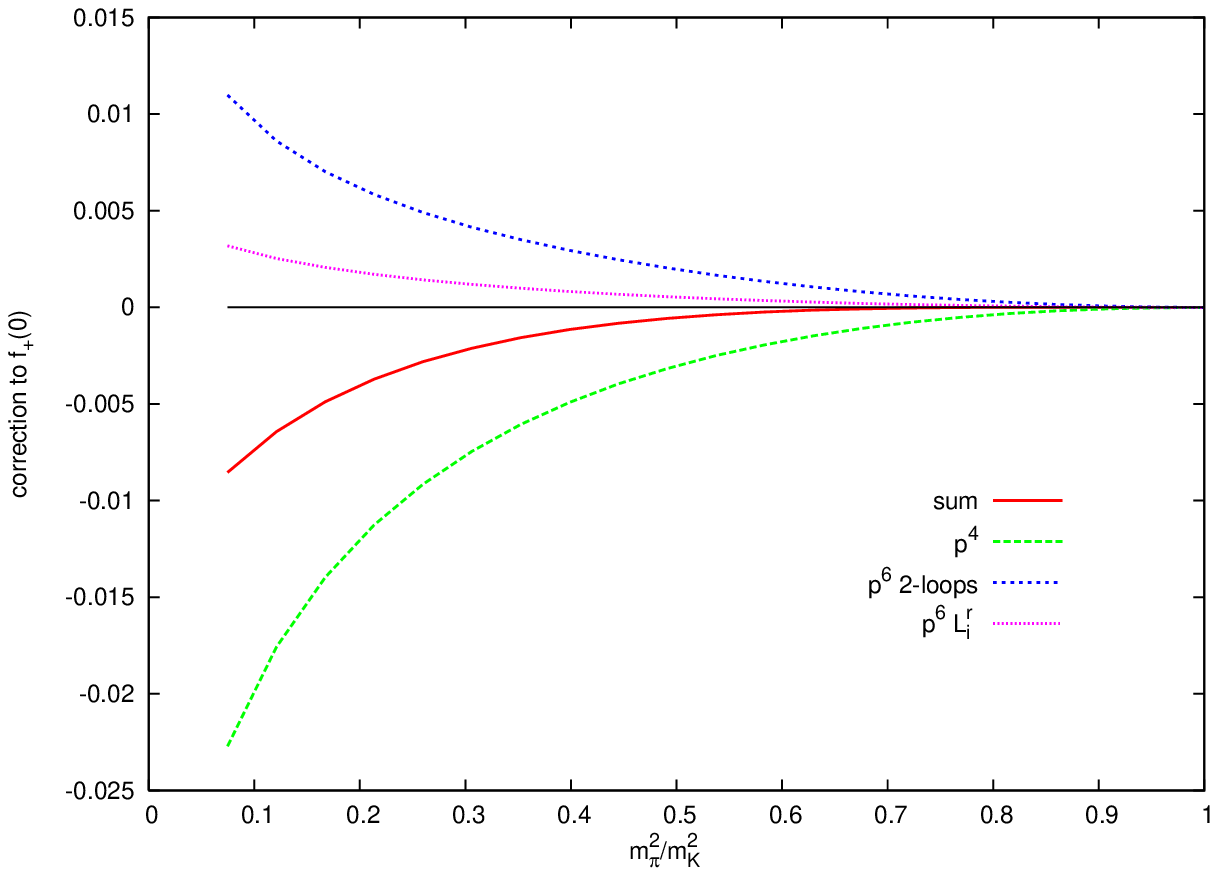}
\centerline{(a)}
\end{minipage}
\begin{minipage}{0.49\textwidth}
\includegraphics[width=\textwidth]{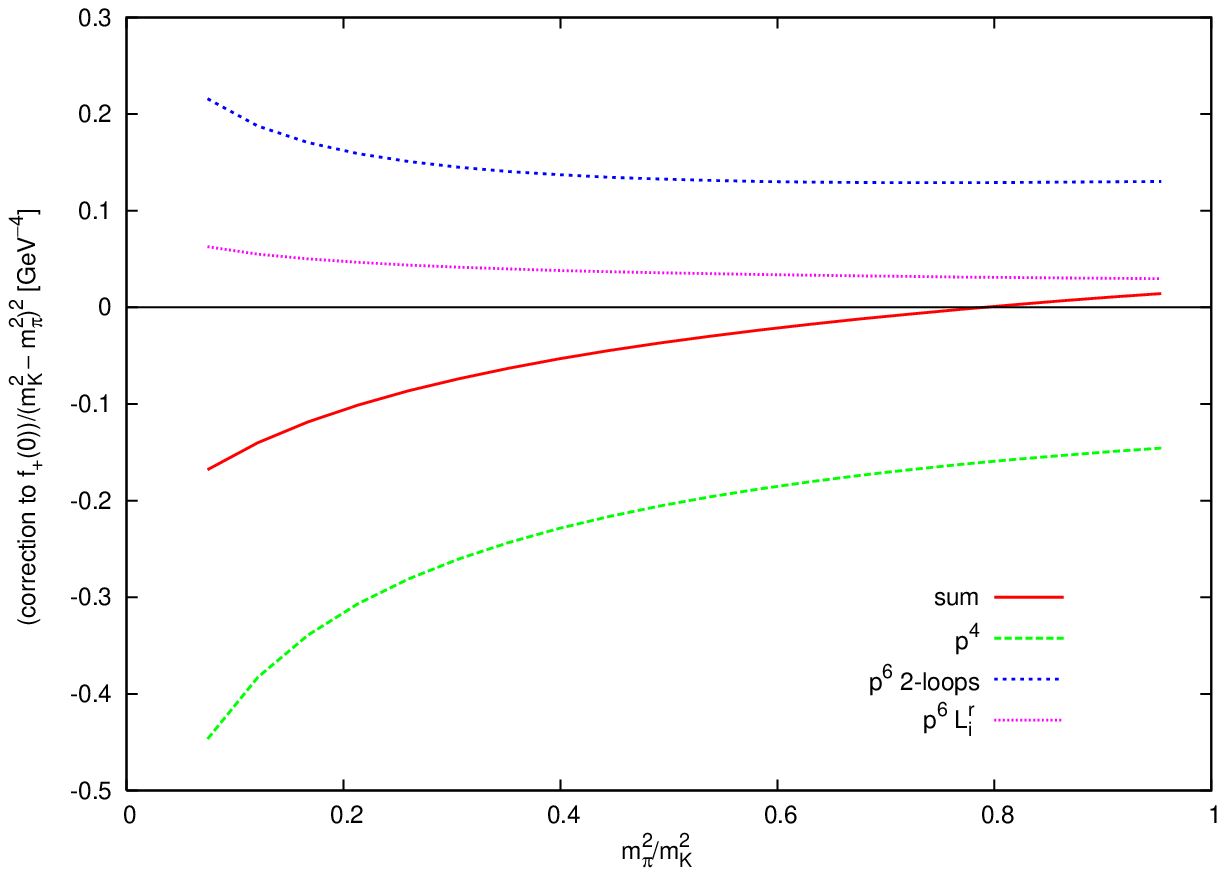}
\centerline{(b)}
\end{minipage}
\caption{For fit 10: (a) $f_+(0)-1$ (b) $(f_+(0)-1)/(m_K^2-m_\pi^2)^2$
as a function of $m_\pi^2$
 with $m_K^2$ fixed.}
\label{figkl3}
\end{figure}

The dependence of $f_+(0)-1$ on the masses is shown in Fig.~\ref{figkl3}.
The left plot shows the three corrections,
order $p^4$, order $p^6$ pure two-loop
part and order $p^6$ the $L_i^r$-dependent part as well as the sum of those
three. It does not include the contribution from the $C_i^r$.
The right plot shows the same contributions divided by 
$\left(m_K^2-m_\pi^2\right)^2$. Both are shown as a function of $m_\pi^2$
with $m_K^2$ fixed. Note that in the right plot
both order $p^6$ corrections are rather flat
except near the physical pion mass.

\section{Even more flavours at NNLO (or PQChPT)}

NLO Partially Quenched ChPT has been studied by many people and found
to be very useful, see \cite{Sharpelectures} and references therein.
The masses and decay constants are known to NNLO for almost all
possible mass combinations. Formulas were kept in terms of the quark-mass
expansion, analogously to (\ref{mpi0}), to avoid the proliferation
in physical masses appearing in this case. The three sea flavour masses
and decay constants can be found in \cite{BDL1,BDL2,BL1} and the two sea
flavour results are in \cite{BL2}. Numerical programs are available
from the authors. The formulas are in the papers but can be downloaded
from \cite{website}. The papers also contain
discussions about how to fit the NNLO LECs.
PQChPT NNLO results for neutral masses are in \cite{BD1}
and
electromagnetism was included
in~\cite{BD2}.

\section{Wishing lists and Conclusions}

The conclusions are several lists. First a general wishing
list and a comment on NNLO fitting:
\begin{itemize}\parskip0cm\itemsep0cm
\item Quark mass dependences \emph{everywhere}
\item A presentation of lattice results at a given quark mass extrapolated
to the continuum and infinite volume. Or, in general, 
results presented in a way that allows us ChPT practitioners
to use your data also later on when other inputs might be changed.

\item More use of the existing NNLO calculations,
if you have any ideas how we can improve the \emph{usability}
  of the existing calculations, please tell us. But remember:
  \begin{itemize}\parskip0cm\itemsep0cm
  \item The number of new parameters at NNLO is very small compared to present
  lattice fitting results.  The new order
  $p^6$ parameters are exactly the same in number as those you add when you add
  an analytic NNLO fitting expression. You therefore do \emph{not}
  loose any predictivity when including the full NNLO result and including
  the known NNLO parts is definitely recommended.
  \item There is a small caveat to this, the order $p^4$ LECs that show up
  in scattering now appear at NNLO also for the masses and decay constants
 \emph{but} these LECs are well known for $n_F=2$, 
  $\bar l_1,\bar l_2$ from Eq. (\ref{valueli}), and $n_F=3$,
  $L_1^r,L_2^r,L_3^r$ from Tab.~\ref{tabfits}. For the partially quenched
  case, there is the unknown extra parameter $\hat L_0^r$, but a large
  $N_c$ estimate gives $\hat L_1^r=\hat L_2^r=0$ and thus $L_0^r = 2 L_1^r$,
   $\hat L_3^r=L_3^r+4 L_1^r$,
  (obtained by inverting (21) of \cite{BDL2}).
  \end{itemize}
\end{itemize}
A more direct wishing list (with input from Berne) for two-flavours
is\\[-0.8cm]
\begin{itemize}\parskip0mm\itemsep0cm
\item $\bar l_3$ and errors.
\item $\bar l_4$: can the lattice check the relation between $F_\pi$ as a
function of the masses and the scalar radius?
\item $\bar l_4$ and \cite{CGL}
Can the lattice check the strong correlation between
the scalar radius and $a^2_0$?
\item Isospin breaking in $\pi\pi$ scattering at $s=m_K^2$, important
for CP-violation phenomenology.
\end{itemize}
A similar list for three flavours
(the different points are rather related)\\[-0.8cm]
\begin{itemize}\parskip0cm\itemsep0cm
\item Large $N_c$ suppressed couplings like $L_4^r$ and $L_6^r$
\item $m_s$ dependence of $F_\pi$ and $m_\pi^2$
\item sigma terms and scalar radii
\item $f_+(0)$ in $K_{\ell3}$ from extrapolations of both $f_+(q^2)$ and
$f_0(q^2)$
\item sigma terms and scalar radii
\end{itemize}
and my final comments
\begin{itemize}\parskip0mm\itemsep0cm
\item Lots of analytical work is done in ChPT at NNLO, please use
(and cite {\Large\Smiley}) it.
\item Use the correct ChPT
  \begin{itemize}\parskip0mm\itemsep0cm
  \item 2-flavour for varying $\hat m$ and possible
    for $N_f=2$ and $N_f=2+1$ at fixed $m_s$ (but have different LECs)
  \item otherwise 3-flavour
  \item the various partially quenched versions
  \end{itemize}
\item Remember at which order in ChPT you compare things: NLO for both
lattice and continuum, NNLO for both lattice and continuum.
Fits put neglected higher order effects into the LECs.
\end{itemize}

And finally, ChPT and LECs played to my great pleasure a large role in many
of the presentations at this conference, as summarized by S.~Necco\cite{Necco}.
I am looking forward to even more
future results from lattice QCD.

\acknowledgments

I want to thank the organizers for a most enjoyable meeting,
S.~D\"urr for useful comments on the manuscript,
my collaborators and the many people at lattice 2007 I discussed
various aspects of lattice QCD and ChPT with.

This work is supported in part by the European Commission RTN network,
Contract MRTN-CT-2006-035482  (FLAVIAnet), 
the European Community-Research Infrastructure
Activity Contract RII3-CT-2004-506078 (HadronPhysics) and
the Swedish Research Council.


\begin{thebibliography}{99}


\bibitem{reviewp6}
J.~Bijnens,
\emph{Prog.\ Part.\ Nucl.\ Phys.}\ { 58} (2007) 521
[hep-ph/0604043].

\bibitem{website} {\tt http://www.thep.lu.se/$\sim$bijnens/chpt.html}

\bibitem{Sharpelectures}
S.~R.~Sharpe,
hep-lat/0607016.

\bibitem{Weinberg0}
S.~Weinberg,
{\em Physica} A { 96} (1979) 327.

\bibitem{GL0}
J.~Gasser and H.~Leutwyler,
{\em Annals Phys.}  { 158} (1984) 142.

\bibitem{GL1}
J.~Gasser and H.~Leutwyler,
{\em Nucl.\ Phys.} B { 250} (1985) 465.

\bibitem{Leutwyler1}
H.~Leutwyler,
{\em Annals Phys.}\  {235} (1994) 165
[hep-ph/9311274].

\bibitem{Descotes-Genon} S.~Descotes-Genon, this conference.

\bibitem{Weinberg1}
S.~Weinberg,
{\em Phys. Rev.}\  {166} (1968) 1568.

\bibitem{BCE1}
J.~Bijnens, G.~Colangelo and G.~Ecker,
{{\em J. High Energy Phys.}} { 9902} (1999) 020
[hep-ph/9902437].

\bibitem{FS2}
H.~W.~Fearing and S.~Scherer,
{\em Phys. Rev.}\ D {53} (1996) 315
[hep-ph/9408346].

\bibitem{BDL1}
J.~Bijnens, N.~Danielsson and T.~A.~L\"ahde,
{\em Phys. Rev.}\ D {70} (2004) 111503
[hep-lat/0406017].

\bibitem{BDL2}
J.~Bijnens, N.~Danielsson and T.~A.~L\"ahde,
{\em Phys. Rev.}\ D {73} (2006) 074509
[hep-lat/0602003]

\bibitem{Haefeli1}
C.~Haefeli, M.~A.~Ivanov, M.~Schmid and G.~Ecker,
arXiv:0705.0576 [hep-ph].

\bibitem{BCE2}
J.~Bijnens, G.~Colangelo and G.~Ecker,
{\em Annals Phys.}\  { 280} (2000) 100
[hep-ph/9907333].

\bibitem{GM}
J.~Gasser and U.~G.~Meissner,
{\em Nucl. Phys.}\ B {357} (1991) 90.

\bibitem{CFU}
G.~Colangelo, M.~Finkemeier and R.~Urech,
{\em Phys. Rev.}\ D {54}, 4403 (1996)
[hep-ph/9604279].

\bibitem{Knechtpipi}
M.~Knecht  {\it et al.}, 
{\em Nucl. Phys.}\ B {457} (1995) 513
[hep-ph/9507319].


\bibitem{BGS}
S.~Bellucci, J.~Gasser and M.~E.~Sainio,
{\em Nucl. Phys.}\ B {423}, 80 (1994)
[Erratum-ibid.\ B {431}, 413 (1994)]
[hep-ph/9401206].

\bibitem{GIS1}
J.~Gasser, M.~A.~Ivanov and M.~E.~Sainio,
{\em Nucl. Phys.}\ B {728} (2005) 31
[hep-ph/0506265].

\bibitem{Burgi1}
U.~Burgi,
%
{\em Phys. Lett.}\ B {377}, 147 (1996)
[hep-ph/9602421].

\bibitem{Burgi2}
U.~Burgi,
{\em Nucl. Phys.}\ B {479} (1996) 392
[hep-ph/9602429].

\bibitem{GIS2}
J.~Gasser, M.~A.~Ivanov and M.~E.~Sainio,
Nucl.\ Phys.\  B {\bf 745} (2006) 84
[hep-ph/0602234].

\bibitem{BCEGS1}
J.~Bijnens {\it et al.}, 
{\em Phys. Lett.}\ B { 374} (1996) 210
[hep-ph/9511397].

\bibitem{BCEGS2}
J.~Bijnens {\it et al.}, 
{\em Nucl. Phys.} B { 508} (1997) 263
[Erratum-ibid.\ B { 517} (1998) 639]
[hep-ph/9707291].

\bibitem{BCT}
J.~Bijnens, G.~Colangelo and P.~Talavera,
{{\em J. High Energy Phys.}} { 9805} (1998) 014
[hep-ph/9805389].

\bibitem{BT1}
J.~Bijnens and P.~Talavera,
{\em Nucl. Phys.}\ B {489}, 387 (1997)
[hep-ph/9610269].

\bibitem{CH}
G.~Colangelo and C.~Haefeli,
{\em Nucl.\ Phys.}\ B { 744}, 14 (2006)
[hep-lat/0602017].

\bibitem{CGL}
G.~Colangelo, J.~Gasser and H.~Leutwyler,
{\em Nucl. Phys.} B { 603} (2001) 125
[hep-ph/0103088].

\bibitem{BT3}
J.~Bijnens and P.~Talavera,
{\em Nucl. Phys.} B { 669} (2003) 341
[hep-ph/0303103].

\bibitem{PS3}
P.~Post and K.~Schilcher,
{\em Eur.\ Phys.\ J.}\ C { 25} (2002) 427
[hep-ph/0112352].

\bibitem{GK1}
E.~Golowich and J.~Kambor,
{\em Nucl. Phys.}\ B {447}, 373 (1995)
[hep-ph/9501318].

\bibitem{ABT1}
G.~Amoros, J.~Bijnens and P.~Talavera,
{\em Nucl. Phys.} B { 568} (2000) 319
[hep-ph/9907264].

\bibitem{DK}
S.~D\"urr and J.~Kambor,
{\em Phys. Rev.}\ D { 61} (2000) 114025
[hep-ph/9907539].

\bibitem{Maltman}
K.~Maltman,
{\em Phys. Rev.}\ D {53} (1996) 2573
[hep-ph/9504404].

\bibitem{Moussallam}
B.~Moussallam,
{\em J. High Energy Phys.} {0008} (2000) 005
[hep-ph/0005245].

\bibitem{Bijnensscalar} J.~Bijnens, unpublished.

\bibitem{GK2}
E.~Golowich and J.~Kambor,
{\em Phys. Rev.}\ D {58}, 036004 (1998)
[hep-ph/9710214].


\bibitem{ABT4}
G.~Amor\'os {\it et al.}, 
{\em Nucl. Phys.} B { 602} (2001) 87
[hep-ph/0101127].

\bibitem{ABT2}
G.~Amor\'os {\it et al.}, 
{\em Phys. Lett.}\ B {480} (2000) 71
[hep-ph/9912398].

\bibitem{ABT3}
G.~Amor\'os {\it et al.}, 
{\em Nucl. Phys.} B { 585} (2000) 293
[Erratum-ibid.\ B { 598} (2001) 665]
[hep-ph/0003258].

\bibitem{BG}
J.~Bijnens and K.~Ghorbani,
{\em Phys.\ Lett.}\  B { 636} (2006) 51
[hep-lat/0602019].


\bibitem{PS1}
P.~Post and K.~Schilcher,
%
{\em Phys. Rev. Lett.}\  {79}, 4088 (1997)
[hep-ph/9701422].

\bibitem{PS2}
P.~Post and K.~Schilcher,
{\em Nucl. Phys.} B { 599} (2001) 30
[hep-ph/0007095].

\bibitem{BT2}
J.~Bijnens and P.~Talavera,
{{\em J. High Energy Phys.}} { 0203} (2002) 046
[hep-ph/0203049].

\bibitem{Geng}
C.~Q.~Geng, I.~L.~Ho and T.~H.~Wu,
{\em Nucl. Phys.}\ B {684} (2004) 281
[hep-ph/0306165].




\bibitem{BD}
J.~Bijnens and P.~Dhonte,
{{\em J. High Energy Phys.}} { 0310} (2003) 061
[hep-ph/0307044].

\bibitem{BDT}
J.~Bijnens, P.~Dhonte and P.~Talavera,
{{\em J. High Energy Phys.}} { 0401} (2004) 050
[hep-ph/0401039].

\bibitem{BDT2}
J.~Bijnens, P.~Dhonte and P.~Talavera,
{\em J. High Energy Phys.} {0405} (2004) 036
[hep-ph/0404150].

\bibitem{Haefeli2}
J.~Gasser, C.~Haefeli, M.~A.~Ivanov and M.~Schmid,
arXiv:0706.0955 [hep-ph].

\bibitem{BG2}
J.~Bijnens and K.~Ghorbani, work in progress.

\bibitem{Kaon07}
J.~Bijnens,
arXiv:0707.0419 [hep-ph].

\bibitem{BL3}
J.~Bijnens and T.A.~L\"ahde, work in progress
\bibitem{Haefeli3}
J.~Gasser, C.~Haefeli, M.~A.~Ivanov and M.~Schmid,
work in progress

\bibitem{Kaiser}
R.~Kaiser,
arXiv:0707.2277 [hep-ph].

\bibitem{KS}
 R.~Kaiser and J.~Schweizer,
{\em J. High Energy Phys.}\ {0606}, 009 (2006)
[hep-ph/0603153].


\bibitem{Ecker1}
G.~Ecker, J.~Gasser, A.~Pich and E.~de Rafael,
{\em Nucl. Phys.} B { 321} (1989) 311.

\bibitem{MHA1}
M.~Knecht and A.~Nyffeler,
{\em Eur.\ Phys.\ J.}\ C { 21} (2001) 659
[hep-ph/0106034];\\

\bibitem{MHA2}
S.~Peris, M.~Perrottet and E.~de Rafael,
{{\em J. High Energy Phys.}} { 9805} (1998) 011
[hep-ph/9805442].

\bibitem{BGLP}
J.~Bijnens {\it et al.}, 
{{\em J. High Energy Phys.}} { 0304} (2003) 055 [hep-ph/0304222].

\bibitem{Reso4}
V.~Cirigliano {\it et al.}, 
{\em Nucl.\ Phys.}\  B { 753} (2006) 139
[hep-ph/0603205].


\bibitem{Pislak1}
S.~Pislak {\it et al.}  [BNL-E865 Collaboration],
{\em Phys. Rev. Lett.}\  { 87} (2001) 221801
[hep-ex/0106071].

\bibitem{Pislak2}
S.~Pislak {\it et al.} [BNL-E865 Collaboration],
{\em Phys. Rev.}\ D { 67} (2003) 072004
[hep-ex/0301040].

\bibitem{Dashen}
J.~Bijnens and J.~Prades,
Nucl.\ Phys.\  B { 490} (1997) 239
[hep-ph/9610360].

\bibitem{Kazimierz}
J. Bijnens, N. Danielsson and T.A. L\"ahde
hep-ph/0701267, to be published in {\em Acta. Phys. Pol.} B.

\bibitem{BDM}
P.~Buettiker, S.~Descotes-Genon and B.~Moussallam,
{\em Eur.\ Phys.\ J.}\ C {33} (2004) 409
[hep-ph/0310283].


\bibitem{BL1}
J.~Bijnens and T.~A.~Lahde,
{\em Phys. Rev.}\ D {71} (2005) 094502
[hep-lat/0501014].

\bibitem{BL2}
J.~Bijnens and T.~A.~Lahde,
{\em Phys. Rev.}\ D {72} (2005) 074502
[hep-lat/0506004].

\bibitem{BD1}
J.~Bijnens and N.~Danielsson,
{\em Phys.\ Rev.}\  D { 74} (2006) 054503
[hep-lat/0606017].

\bibitem{BD2}
J.~Bijnens and N.~Danielsson,
{\em Phys.\ Rev.}\  D { 75} (2007) 014505
[hep-lat/0610127].

\bibitem{Necco}
S.~Necco, this conference.
\end{thebibliography}
\end{document}